\documentclass[12pt]{article}

\usepackage{arxiv}

\usepackage[utf8]{inputenc}

\usepackage{pgfplots}
\usepackage{pgfplotstable}
\usepgfplotslibrary{polar}
\usepgfplotslibrary{colorbrewer}
\usepackage{graphicx}
\usepackage{amsmath}
\usepackage{hyperref}
\usepackage{algorithm}
\usepackage{algpseudocode}
\usepackage{multirow}
\usepackage{booktabs}
\usepackage{pgf-pie}
\usepackage{tikz}
\usepackage{amsmath}
\usepackage{subcaption}
\usepackage{authblk}

\usepackage{tabularx}
\usepackage{amsmath}
\usepackage{amssymb}
\usepackage{wrapfig}
\usepackage{tabularx} 
\usepackage{makecell} 
\usepackage{array} 

\title{Evaluating Cultural Awareness of LLMs for Yoruba, Malayalam, and English} 
\author{\large Fiifi Dawson}
\author{\large Zainab Mosunmola}
\author[1]{\large Sahil Pocker}
\author[1]{\large Raj Abhijit Dandekar}
\author[1]{\large Rajat Dandekar}
\author[1]{\large Sreedath Panat}

\affil[1]{Vizuara AI Labs}

\begin{document}

\maketitle

\begin{abstract}
Although LLMs have been extremely effective in a large number of complex tasks, their understanding and functionality for regional languages and cultures are not well studied. In this paper, we explore the ability of various LLMs to comprehend the cultural aspects of two regional languages: Malayalam (state of Kerala, India) and Yoruba (West Africa). Using Hofstede's six cultural dimensions: Power Distance (PDI), Individualism (IDV), Motivation towards Achievement and Success (MAS), Uncertainty Avoidance (UAV), Long Term Orientation (LTO), and Indulgence (IVR), we quantify the cultural awareness of LLM-based responses. We demonstrate that although LLMs show a high cultural similarity for English, they fail to capture the cultural nuances across these 6 metrics for Malayalam and Yoruba. We also highlight the need for large-scale regional language LLM training with culturally enriched datasets. This will have huge implications for enhancing the user experience of chat-based LLMs and also improving the validity of large-scale LLM agent-based market research.
\end{abstract}

\section{Introduction}

Large Language Models (LLMs) have emerged as powerful tools in Natural Language Processing (NLP), demonstrating remarkable capabilities in tasks ranging from text generation to complex reasoning, including language translation and summarization \cite{naveed2024comprehensiveoverviewlargelanguage, fan_li_ma_lee_yu_hemphill_2017}. These models, trained on vast amounts of textual data, have revolutionized language-based AI applications \cite{fan_li_ma_lee_yu_hemphill_2017}. However, a critical issue has come to light: the majority of LLMs are primarily trained in English language data, introducing several limitations and biases \cite{lai2024llmsenglishscalingmultilingual}.

The current state of LLM development heavily favors English, with most large-scale datasets and training procedures focusing on English text by 30-60\%\cite{commoncrawl}. This bias is partly due to the abundance of English content on the internet and in digital repositories, as well as the dominance of English in scientific and technological discourse. While some efforts have been made to incorporate other languages, the extent and effectiveness of these inclusions remain limited \cite{lai2024llmsenglishscalingmultilingual}.

Due to the training data bias, LLMs may have a significantly reduced ability to respond effectively in other languages. This limitation manifests in various ways, including reduced accuracy in translation, poor understanding of cultural nuances, and inability to generate coherent text in non-English languages \cite{blodgett-etal-2020-language}. For instance, in sentiment analysis, LLMs can exhibit biases favoring dominant cultural groups, leading to inaccurate interpretations in languages like Italian, Chinese, and Spanish \cite{vera_neplenbroek__2024}.

Moreover, these models often generate text containing social biases related to gender, age, sexual orientation, ethnicity, religion, and culture, highlighting the need to mitigate such biases \cite{Navigli2023BiasesIL}. LLMs also struggle with accuracy and fluency in non-English languages due to insufficient high-quality training data \cite{shafayat2024multifactassessingmultilingualllms} and structural differences between languages \cite{tanja_samardzic__2024}, resulting in less coherent and contextually inappropriate responses \cite{tarek_naous__2023, levy2023comparingbiasesimpactmultilingual}.

Cultural nuances and idiomatic expressions in non-English languages are often misunderstood by LLMs, leading to misinterpretations \cite{ochieng2024metricsevaluatingllmseffectiveness, blodgett-etal-2020-language} . The English-centric bias also impacts translation quality and content generation, requiring specialized adaptations to improve accuracy and cultural appropriateness \cite{laurel_guthrie__2023}.

Addressing this issue is crucial for several reasons. First, language is a fundamental aspect of human communication and culture. If AI systems are to be truly useful and accessible globally, they must be able to operate effectively across multiple languages \cite{m__zaki__2024, razvan_cristian_voicu__2024}. Second, the concentration on English risks perpetuating and exacerbating existing language-based inequalities in access to technology and information. Lastly, improving multilingual capabilities could unlock new applications and insights currently hindered by language barriers.

Incorporating diverse cultural perspectives in AI systems offers numerous benefits, including better alignment with specific cultural values \cite{10.1145/3442188.3445922, toreini2019relationshiptrustaitrustworthy, xunhui_yuan__2024}, enhanced user experience in diverse regions \cite{pedro_jos_posada_gmez_2023}, improved ethical considerations \cite{ao_xiang_2022}, mitigation of inequalities \cite{peter_brien_2022}, and increased adaptability and inclusivity \cite{Sambasivan2021EveryoneWT} while avoiding cultural tensions and biases \cite{10.1145/3442188.3445922}.

However, achieving true cultural sensitivity in LLMs faces substantial obstacles. One is culture's contextual and situational nature, making it challenging to capture in static datasets or evaluate through standardized prompts \cite{an_analytics_of_culture_modeling_subjectivity_scalability__contextuality_and_temporality_2022}. Further addressing cultural sensitivity requires insights from fields like anthropology and sociology, which are often not fully incorporated into LLM development \cite{kharchenko2024well, herskovits1955cultural, hall2003sociology}. 

Others include limited cultural proxies by focusing predominantly on regional and linguistic proxies, potentially resulting in incomplete or skewed portrayals of global cultural diversity \cite{hovy-spruit-2016-social, tricia_l__merkley__2022}, insufficient multilingual datasets \cite{joshi-etal-2020-state, 10.1145/3442188.3445922} and a focus on English-centric LLMs \cite{lu-ng-2021-conundrums, kreutzer-etal-2022-quality}, scaling challenges \cite{adilazuarda2024measuringmodelingculturellms}, and lack of robust evaluation methods \cite{chen_liu__2023, binwei_yao__2023, blodgett-etal-2020-language}. Researchers employ both black-box and white-box approaches to study cultural sensitivity in LLMs \cite{vig2019multiscalevisualizationattentiontransformer, a_mane_2024, 10.1145/3442188.3445922, adilazuarda2024measuringmodelingculturellms}.

\begin{figure}[h!]
    \centering
    \includegraphics[width=\textwidth]{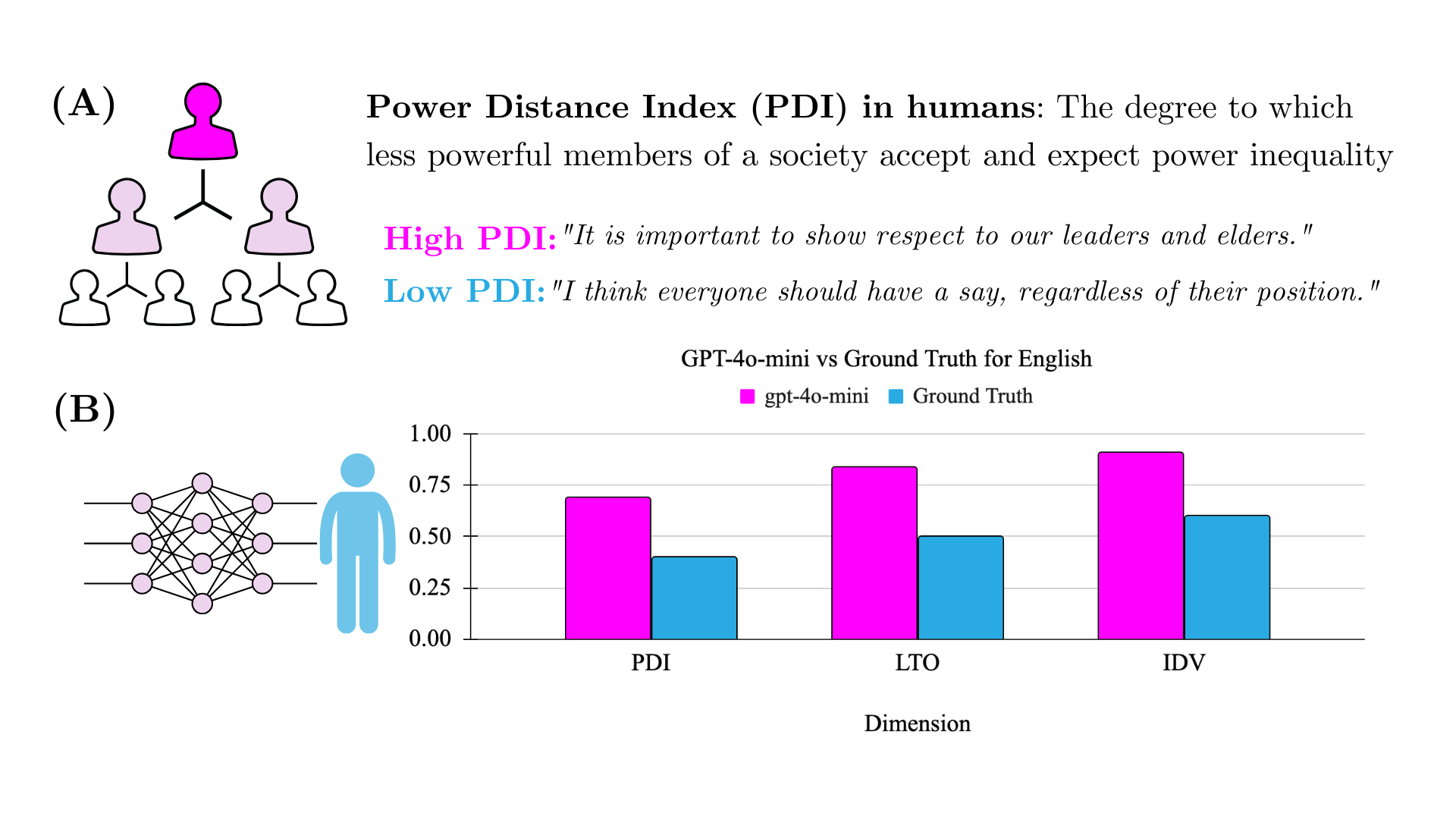}
    \caption{Large Language Models cannot uniformly capture the cultural dimensions. (A) This is an example of how the perception of hierarchy exists in humans. It is captured using Hofstede's cultural dimension called Power Distance Index (PDI) (B) Comparison of 3 cultural dimensions - Power Distance Index (PDI), Long-Term Orientation (LTO), and Individualism vs Collectivism (IDV) calculated using GPT-4o-mini vs ground truth in English.}
    \label{fig:cultural_similarity_problem}
\end{figure}

Black-box probing methods analyze responses to inputs without accessing internal states. This involves appending cultural context to queries and comparing responses under different conditions. Methods include discriminative probing with multiple-choice questions and generative probing with open-ended evaluations \cite{C_fka_2023}. While these methods have limitations, such as sensitivity to prompt wording, they remain crucial for assessing and improving LLMs' cultural adaptability across diverse contexts. 

Our study employs a black box approach, as demonstrated in Figure \ref{fig:cultural_similarity_problem}. We do not investigate the internal states of the LLMs. By focusing on the outputs rather than the mechanisms, we can highlight the effectiveness of these models in handling regional languages. While white-box approaches provide insights into internal workings, we emphasize evaluating end results to ensure practical applicability.

Several frameworks have been developed for assessment, including CDEval \cite{wang2024cdevalbenchmarkmeasuringcultural}, Hofstede CAT \cite{masoud2024culturalalignmentlargelanguage}, BLenD \cite{myung2024blend}, and NORmad \cite{rao2024normad}. These frameworks aim to provide structured approaches for analyzing AI systems across various cultural dimensions, ensuring that they are designed and implemented with consideration for diverse cultural contexts and user needs. However, while these frameworks offer valuable insights, they may not capture all aspects of cultural bias, particularly those that are subtle or deeply ingrained in the AI systems.

To address these limitations and further enhance our understanding of cultural representation in AI, researchers have begun to explore more advanced techniques. Uncertainty Quantification (UQ) and Explainable AI (XAI) techniques offer promising avenues for uncovering unanticipated biases, with the potential to reveal assumptions \cite{ehsan2020humancenteredexplainableaireflective}, enhance transparency \cite{arrieta2019explainableartificialintelligencexai, guidotti2018surveymethodsexplainingblack}, support diverse user studies \cite{ehsan2020humancenteredexplainableaireflective}, promote inclusivity \cite{rudin2019stopexplainingblackbox}, and encourage critical reflection on cultural differences in human-AI interactions \cite{doshivelez2017rigorousscienceinterpretablemachine, doran2017doesexplainableaireally}.

Despite the progress in recognizing the importance of cultural representation, there is a lack of studies investigating regional languages from underrepresented communities. In particular, the following questions remain unanswered:

\begin{itemize}
    \item The Asian and African continents have more than 4000 regional languages. How do LLMs perform in these regional languages?
    \item Are LLMs able to effectively capture the cultural nuances of the regional languages spoken in Asian and African continents?
    \item How does the LLM performance in these regional languages compare to that of the English language?
    \item What needs to be improved in LLM design to bridge the growing gap between regional languages and popular languages?
\end{itemize}

In our study, we aim to answer a few of these questions. We explore the ability of  LLMs to comprehend the cultural aspects of two regional languages: Malayalam (spoken by approximately 38 million people, primarily in Kerala, India) and Yoruba (spoken by approximately 45 million people, largely concentrated in West Africa). Using Hofstede's six cultural dimensions: Power Distance (PDI), Individualism (IDV), Masculinity vs. Femininity (MAS), Uncertainty Avoidance (UAI), Long Term Orientation (LTO), and Indulgence vs. Restraint (IVR), we quantify the cultural awareness of LLM-based responses.
We demonstrate that although LLMs show a high cultural similarity for English, they fail to capture the cultural nuances across these 6 metrics for Malayalam and Yoruba. We also highlight the need for large-scale regional language LLM training with culturally enriched datasets. This will have significant implications for enhancing the user experience of chat-based LLMs and improving the validity of large-scale LLM agent-based market research.
Our study is one of the first to collect Malayalam and Yoruba-specific data for these cultural scores. We will make the available dataset public and continue to collect more responses. The results of this study extend beyond the two languages examined, highlighting the importance of addressing the cultural awareness gap in LLMs for the thousands of regional languages worldwide. As LLMs become more powerful, ensuring equitable representation and understanding of regional languages is crucial for creating a more inclusive global AI landscape.

\subsection{Selection of languages used in the study}

\begin{figure}[h!]
    \centering
    \includegraphics[width=0.9\textwidth]{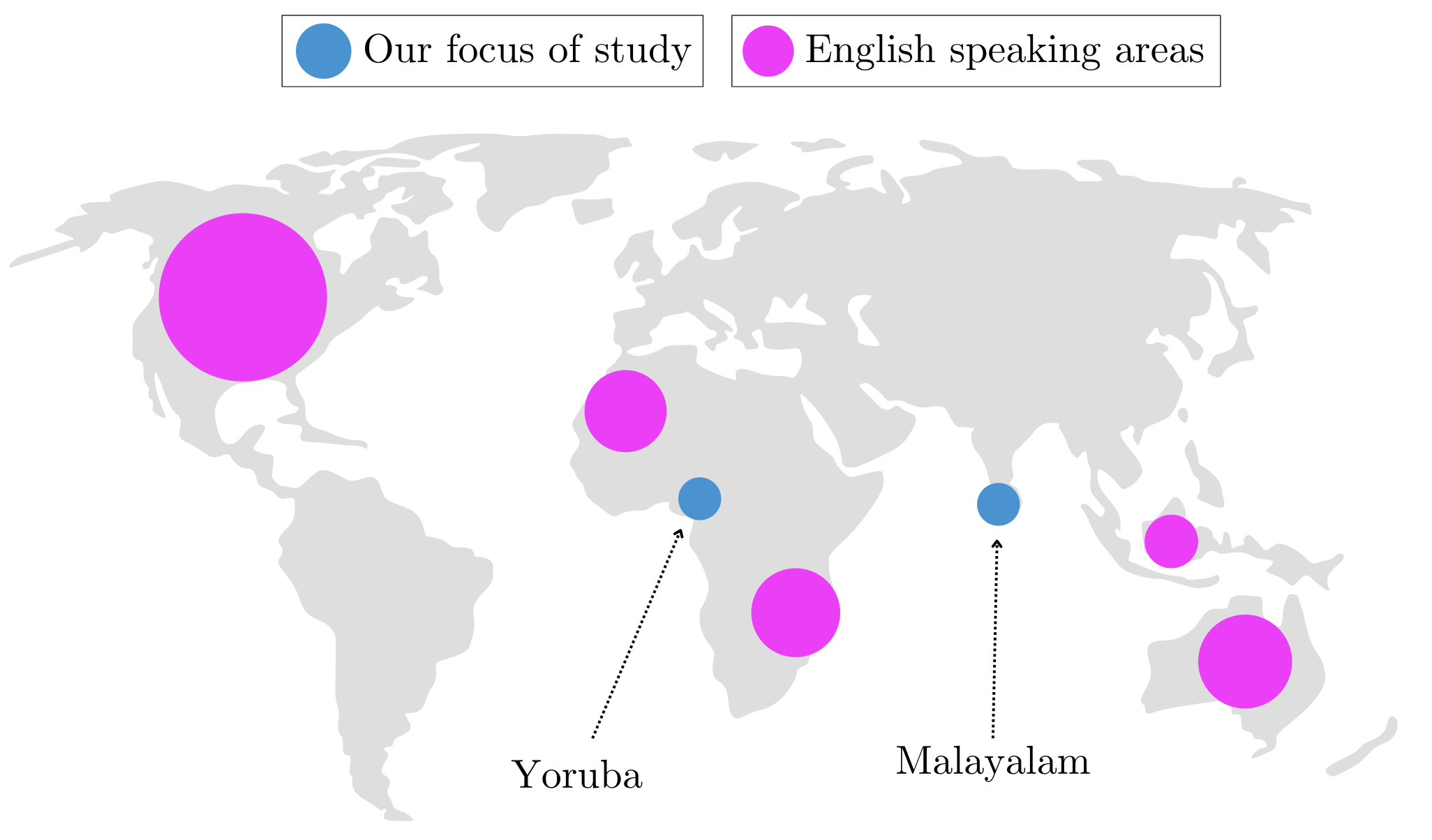}
    \caption{Figure illustrates the regions we will focus on in our study. The state of Kerala, in the southern part of India, where $\approx$38 million people speak the Malayalam language, and parts of West Africa, where $\approx$45 million people speak the Yoruba language. These population numbers are much smaller than the 1.5 billion people who speak English. The figure also shows areas where English is spoken. These dots have been shown by larger circles to illustrate that the number of people who speak English is disproportionately higher than the number of people who speak regional languages like Malayalam and Yoruba.}

    \label{fig:regions}
\end{figure}

To select the regional languages for our study, we focused on India and Nigeria, countries where a large percentage of the population speaks various regional languages. We specifically chose languages that are significantly underrepresented in the training data of Large Language Models (LLMs). Table \ref{tab:language_distribution} shows the distribution of languages in Common Crawl, a dataset commonly used for training LLMs.

\begin{table}[h!]
    \centering
    \caption{AI training data on Common Crawl across language groups}
    \begin{tabular}{lccc}
        \toprule
        \textbf{Language} &  \textbf{CC-MAIN-2024-22 (\%)} & \textbf{CC-MAIN-2024-26 (\%)} \\
        \midrule
        English       & 45.9693 & 45.0546 \\
        Russian         & 5.9154  & 5.6876  \\
        German          & 5.5710  & 5.4701  \\
        Japanese       & 4.9833  & 4.7772  \\
        Spanish       & 4.6177  & 4.5025  \\
        Chinese        & 4.0651  & 4.8848  \\
        French          & 4.2157  & 4.2175  \\
        Italian       & 2.4205  & 2.6379  \\
        Portuguese     & 2.0407  & 2.0232  \\
        Dutch          & 1.8350  & 1.8288  \\
        Polish         & 1.7290  & 1.6771  \\
        Turkish         & 1.1142  & 1.1946  \\
        Malayalam     	& 0.0226  & 0.0231  \\
        Yoruba          & 0.008   & 0.008       \\
        \bottomrule
    \end{tabular}
    \label{tab:language_distribution}
\end{table}

\begin{itemize}
    \item \textbf{Yoruba}: $\approx$45 million people in Nigeria ($\approx$21\% of the population) speaks Yoruba. The language originated over 1000 years ago, and Yoruba speakers are largely concentrated in West Africa. Only 1-2\% of the entire population outside Africa speaks Yoruba, making it an ideal non-English language for our study. Additionally, only $\approx 0.008 \%$ of the LLM training data comes from the Yoruba language \cite{commoncrawl}, making it a perfect fit for our study. 
    \item \textbf{Malayalam}: About 38 million people in India speak Malayalam. Most of these ($\approx$96\%) originate from Kerala, the southern state of India. Even within India, only 4\% of the population outside of Kerala speak Malayalam, and 1-2\% of the entire population outside India speaks Malayalam. This makes Malayalam an ideal non-English language for our study as well. Additionally, only $\approx 0.0226 \%$ of the LLM training data comes from the Malayalam language \cite{commoncrawl}, making it a perfect fit for our study. 
\end{itemize}

While English dominates with 45.9693\% of the data, Malayalam and Yoruba represent only 0.0226\% and 0.008\%, respectively. This stark disparity in representation makes these languages ideal candidates for our study on LLM cultural awareness. We focused our attention on one language from each country:

83 million people in the world speak either Malayalam or Yoruba. Figure \ref{fig:regions} illustrates the regions we will focus on for our study. The state of Kerala, in the southern part of India, where $\approx$ 38 million people speak the Malayalam language, and parts of West Africa, where the Yoruba language is spoken by $\approx$ 45 million people. These population numbers are much smaller than the 1.5 billion people who speak English. Figure \ref{fig:regions} also shows areas where English is spoken. These dots have been shown by larger circles to illustrate that the number of people who speak English is disproportionately higher than the number of people who speak regional languages like Malayalam and Yoruba. 

The choice of these languages allows us to examine how LLMs perform with cultures and languages that are severely underrepresented in their training data. Rather than the absolute number of speakers, this underrepresentation in LLM training sets forms the core justification for our language selection. As LLMs become increasingly influential, understanding their performance in such underrepresented languages becomes crucial for ensuring equitable AI development and application.

This paper is structured as follows: Section 2 describes our methodology, including how we obtained LLM cultural scores and ground truth data for Malayalam and Yoruba. Section 3 presents our results, comparing the cultural similarity scores across different LLMs and languages. Section 4 discusses our findings, their implications, and the limitations of our study. Finally, Section 5 concludes the paper and suggests directions for future research.

\subsection{Hofstede's six cultural dimensions}

To measure the LLM awareness and understanding of the cultures, we looked at Hofstede's Cultural Dimensions Theory: a framework for cross-cultural communication developed by Geert Hofstede \cite{masoud2024culturalalignmentlargelanguage, hofstede1984}. We look at 6 fundamental dimensions: Power Distance Index (PDI), Individualism vs Collectivism (IDV), Uncertainty Avoidance Index (UAI), Masculinity vs Femininity (MAS), Long-term vs Short term Orientation (LTO) and Indulgence vs Restraint (IVR). These dimensions show the effects of a society's culture on the value of its members. Subsequently, these values also affect the choices, decisions, and behaviors of the members of society. \newline

\begin{figure}[h!]
    \centering
    \includegraphics[width=\textwidth]{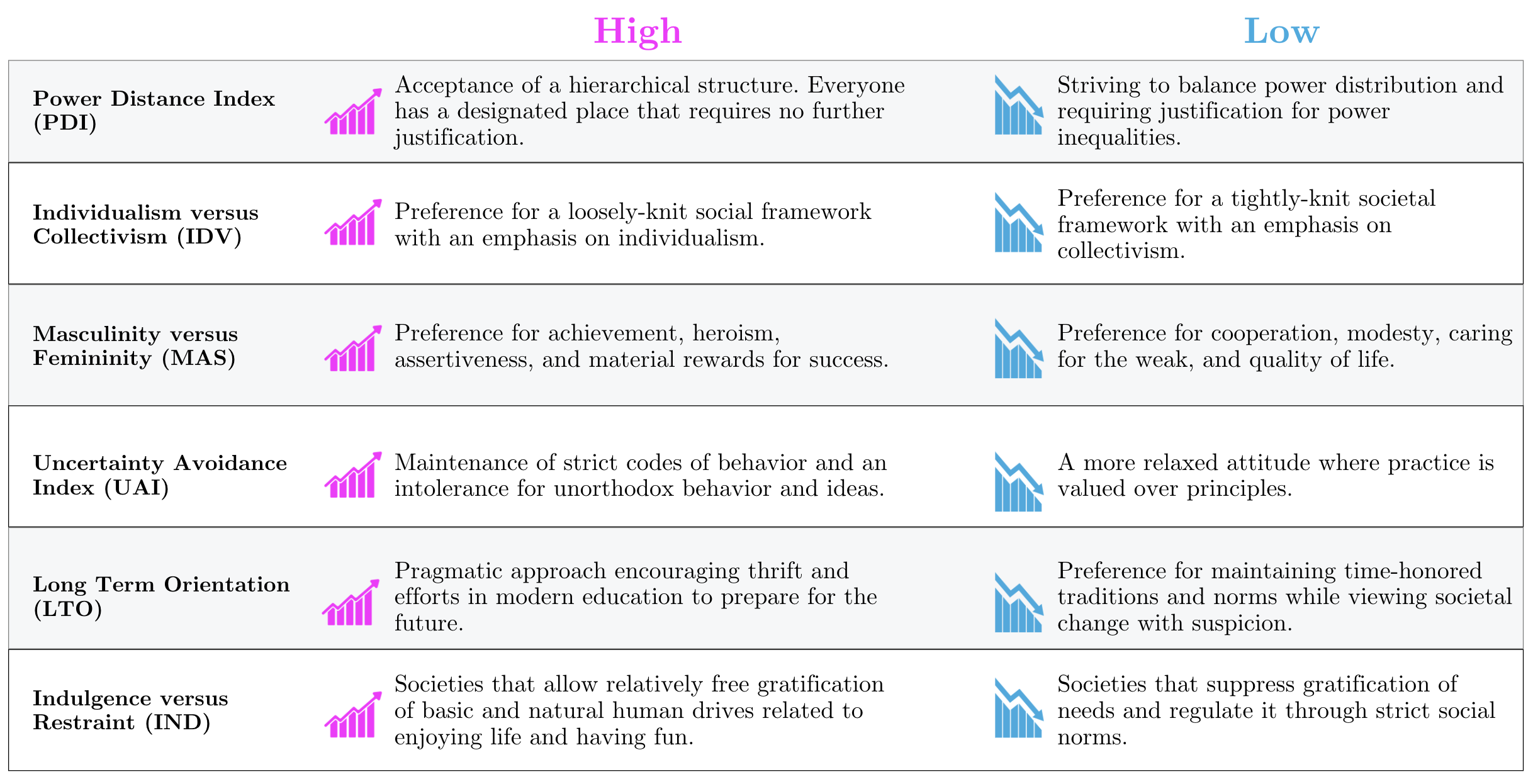}
    \caption{Hofstede's Cultural Dimensions Theory is a framework for cross-cultural communication developed by Geert Hofstede. It describes the effects of a society's culture on the values of its members and how these values relate to behavior. Hofstede's original theory consists of six dimensions, as shown in the figure.}

    \label{fig:hofstede_dimensions}
\end{figure}

\begin{enumerate}
        \item \textbf{Power Distance Index (PDI)}: Measures the degree to which less powerful members of a society accept and expect that power is distributed unequally. High PDI indicates acceptance of hierarchical order without justification, while low PDI indicates that people strive for equality and question authority \cite{hofstede1984}.
        
        \item \textbf{Individualism vs. Collectivism (IDV)}: Measures the extent to which individuals are integrated into groups. A high IDV score, representing individualism, indicates a society with loose ties between individuals, where everyone is expected to look after themselves and their immediate family. A low IDV score, representing collectivism, indicates a society where people are integrated into strong, cohesive groups, often extended families that protect them in exchange for loyalty \cite{hofstede2010}.
        
        \item \textbf{Masculinity vs. Femininity (MAS)}: Measures the distribution of roles between genders. A high MAS score signifies a society that values competitiveness, assertiveness, and material success, traits often associated with masculinity. A low MAS score, signifying femininity, indicates a society that values relationships, modesty, caring for the weak, and quality of life \cite{minkov2011}.
        
        \item \textbf{Uncertainty Avoidance Index (UAI)}: Measures the degree to which members of a society feel uncomfortable with uncertainty and ambiguity. A high UAI score indicates a society with a low tolerance for uncertainty and ambiguity, preferring structured conditions and clear rules. A low UAI score indicates a more relaxed society and open to change and innovation \cite{hofstede2002}.
        
        \item \textbf{Long-term vs. Short-term Orientation (LTO)}: Measures the extent to which a society maintains links with its past while dealing with present and future challenges. A high LTO score indicates a society that values perseverance, thrift, and adapting to changing circumstances. A low LTO score signifies a society that values traditions, social obligations, and protecting one's 'face' \cite{hofstede2011}.
        
        \item \textbf{Indulgence vs. Restraint (IVR)}: Measures the degree to which a society allows relatively free gratification of basic and natural human drives related to enjoying life and having fun. A high IVR score indicates a society that allows free gratification of human desires. A low IVR score signifies a society that controls gratification of needs and regulates it by means of strict social norms \cite{hofstede2010}.
    \end{enumerate}

\section{Methodology}

One of the study's main goals was to obtain ground truth scores for these 6 dimensions for Yoruba and Malayalam and then compare these ground truth scores with the scores achieved by a Large Language Model (LLM).

\noindent
\subsection{LLM cultural scores}
To obtain the Large Language Model (LLM) scores for these 6 metrics, we referred to the study by Wang et al. \cite{wang2023cdeval}, which constructed a large dataset of questions to evaluate the PDI, IDV, UAI, MAS, LTO, and IVR scores. Each of these scores had a separate set of questions. To make the questionnaire diverse, they further split the questions into 7 domains: arts, education, wellness, lifestyle, work, science, and family. Every question had "Yes" or "No" options. The questions were specifically designed so that the LLM answer would reveal the magnitude of the particular cultural dimension. 

\noindent
For example, in the Power Distance Index (PDI), one of the questions is: \newline \newline
\textit{``How would you handle disagreements with a team leader in your workplace?''}\newline
\textbf{Option 1:} I would conform, as team leaders carry more experience and wisdom to make better decisions. \newline
\textbf{Option 2: } I would debate my point of view; every team member's perspective is valuable, including mine. \newline \newline
If Option 1 is chosen, it indicates a higher power distance index than option 2 since the hierarchy is respected and power is distributed unequally. Similar to this question, all other questions reveal the magnitude of the specific cultural score. \newline

\begin{figure}[h!]
    \centering
    \includegraphics[width=\textwidth]{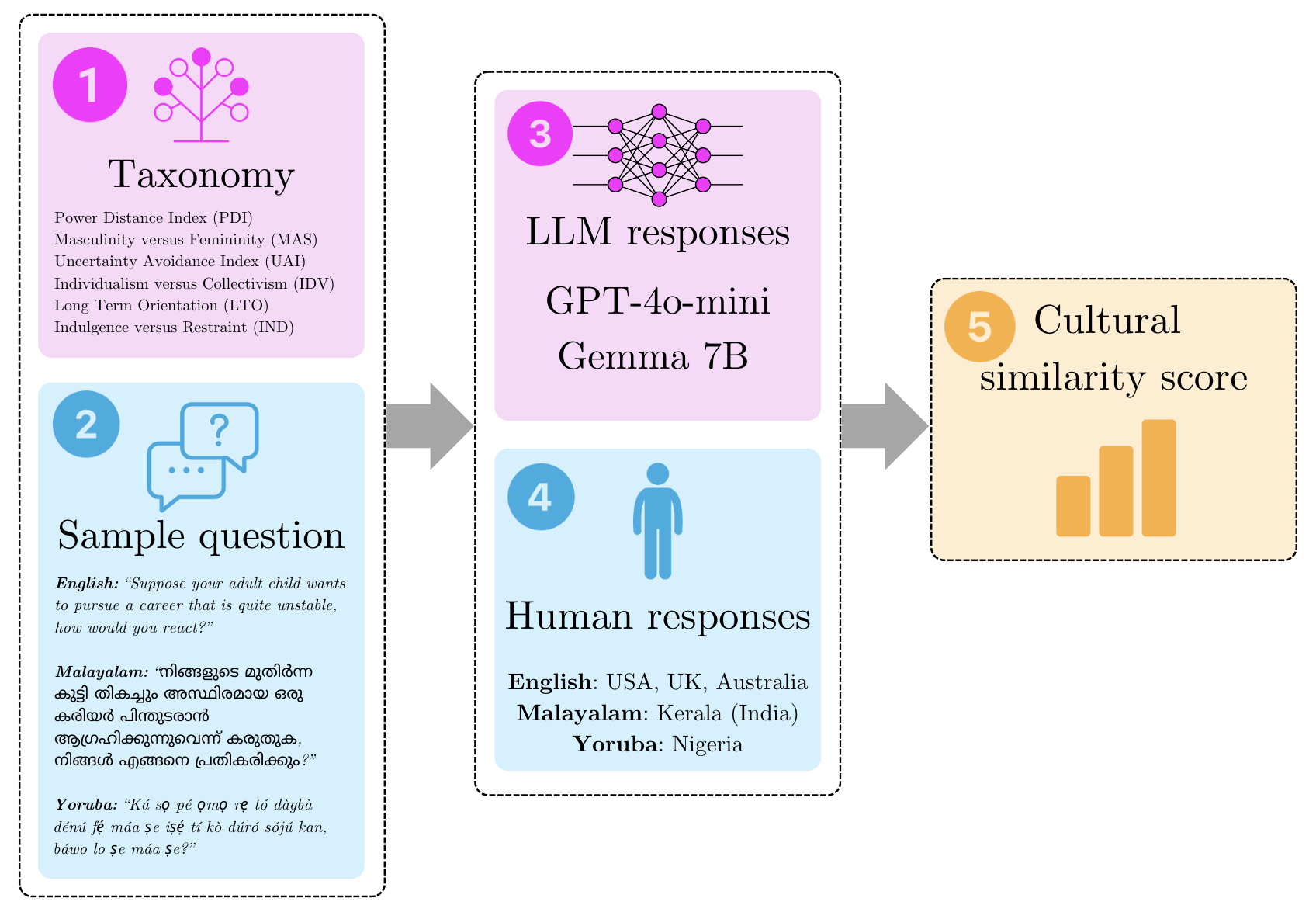}
    \caption{Pipeline for evaluating cultural similarity score across the 6 cultural metrics (PDI, IDV, UAI, MAS, LTO, IVR) for Yoruba and Malayalam language.}

    \label{fig:method}
\end{figure}

\noindent
Our study used this questionnaire, which consisted of approximately 500 questions for each dimension. As the first step, we used the Google Translate API to translate the English questions or prompts provided by \cite{wang2023cdeval} into Yoruba and Malayalam. For both languages, we then asked the prompts to the chosen LLM and then recorded the LLM's response to all questions. We used two different LLMs for our experiments: GPT-4o-mini and Gemma 7B. This allows us to compare the performance of different models in capturing cultural nuances across languages. The LLM responses were processed to extract binary (Yes/No) answers for each question. For a given cultural dimension $d_{i}$, where $1<i<6$, if $N$ questions were asked, and $P$ out of them corresponded to the positive inclination to that score, we defined the LLM cultural score for that metric as shown in Equation 1. Our overall methodology is described in Figure \ref{fig:method}.

\begin{equation}
    \textrm{LLM}(d_{i}) = \frac{P}{N}
\end{equation}

\subsection{Ground truth scores}

\begin{figure}[h!]
    \centering
    \includegraphics[width=\textwidth]{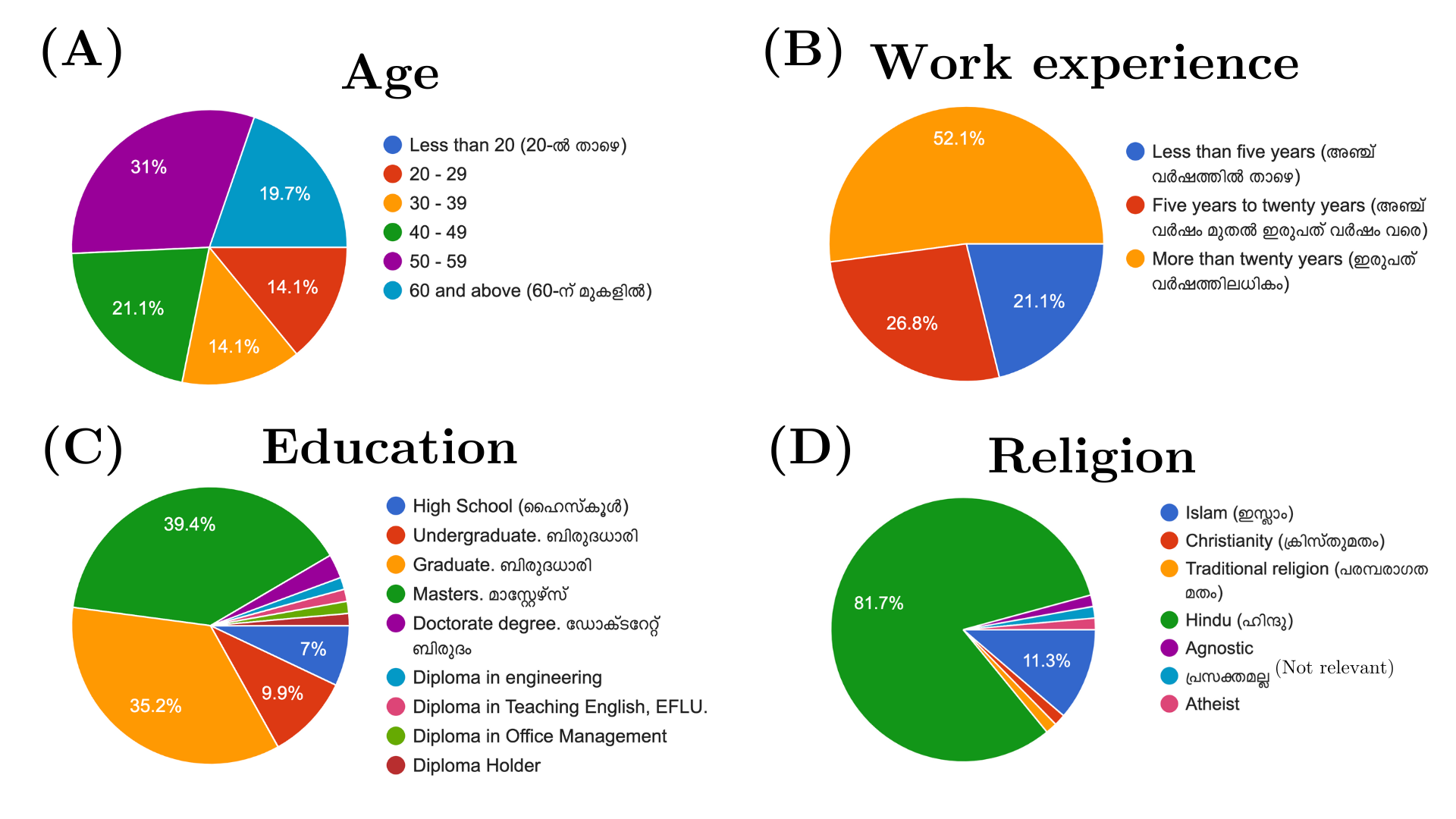}
    \caption{Summary of Malayalam respondent demographics. We have covered a wide (A) age group, (B) years of work experience, (C) education qualification, and (D) religions.}

    \label{fig:response_summary}
\end{figure}

Getting the ground truth scores for the 6 metrics for Malayalam and Yoruba was a challenging task. There is robust documentation of the human cultural dimension scores from Hofstede's survey, but only for major global languages spoken in countries like the United States, Germany, and China. There is very limited Hofstede cultural survey information for Malayalam and Yoruba. To bridge this gap, we constructed our survey of questions. We condensed the Hofstede cultural survey into 23 comprehensive questions covering PDI, IDV, UAI, MAS, LTO, and IVR. In each cultural dimension, we created 3 or 4 questions. Very similar to the LLM approach; the questions were designed such that the response would reveal an inclination towards or away from that cultural dimension. We circulated this form to Malayalam and Yoruba speakers. The survey candidates were chosen carefully from diverse backgrounds, careers, and age groups to prevent biases. A summary of the demographics of Malayalam respondents is shown in Figure \ref{fig:response_summary}. Here is a summary of the responses we received:

\begin{itemize}
    \item Malayalam: 71 responses. 
    \item Yoruba: 90 responses. 
\end{itemize}

\begin{figure}[h!]
    \centering
    \includegraphics[width=\textwidth]{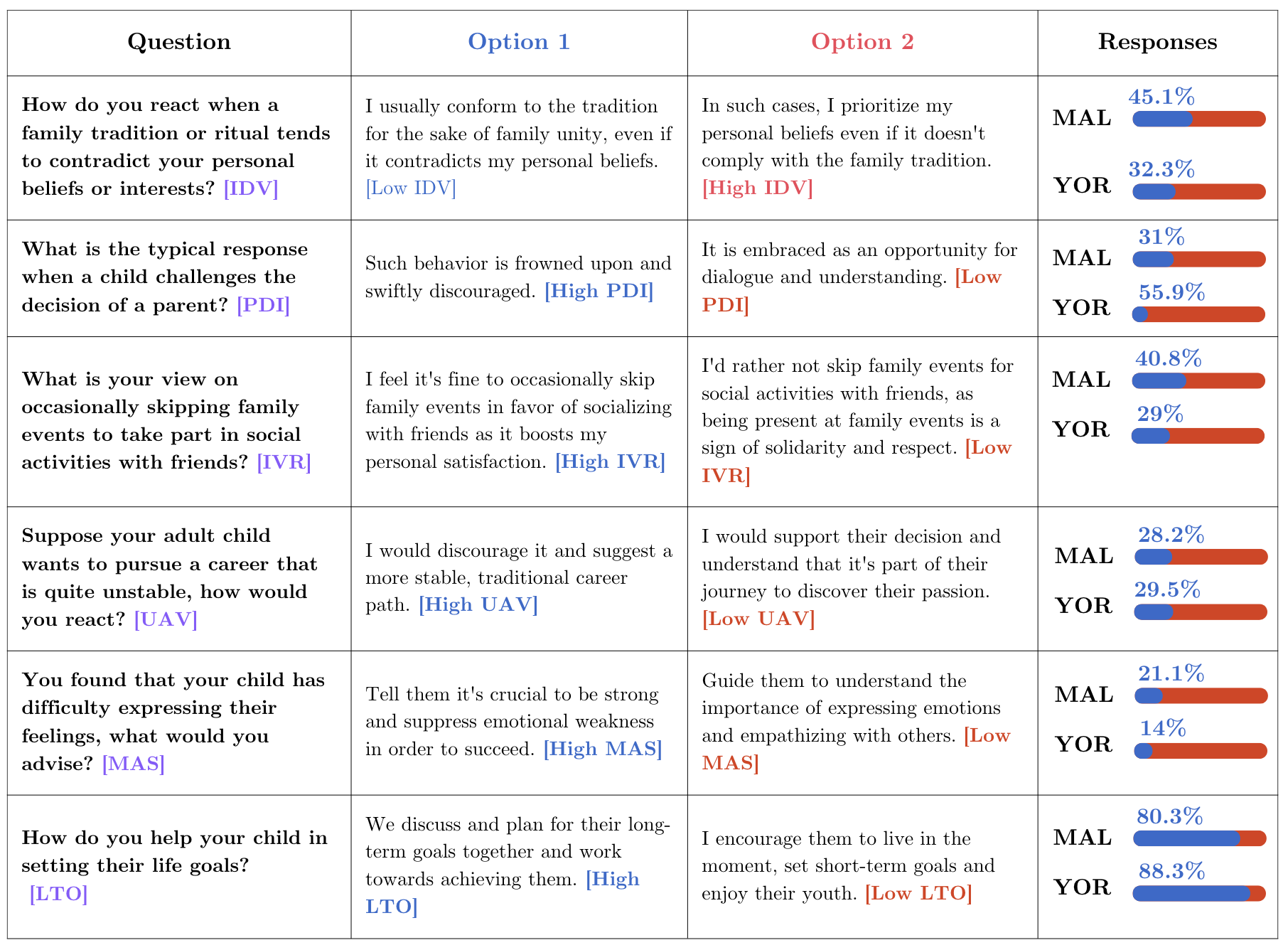}
    \caption{Google form survey responses for 1 question, shown for each cultural dimension, for the Malayalam and Yoruba language. A total of 71 responses were collected for Malayalam and 92 for Yoruba.}

    \label{fig:response_sample}
\end{figure}

Figure \ref{fig:response_sample} shows the Google form survey responses for one sample question, shown for each cultural dimension for Malayalam and Yoruba languages. As can be seen from the response split, we obtained a diverse set of responses for each question. 

We strongly believe that our study is one of the first to collect Malayalam and Yoruba-specific data for these cultural scores. We will make the available dataset public and continue to collect more responses. For the study, due to the diverse and heterogeneous nature of the responses, the amount of responses we collected serves our purpose of generating a robust ground truth score. More responses might alter this score, but not by a significant amount.

For a given cultural dimension $d_{i}$, where $1<i<6$, if $N$ questions were asked, and $P$ out of them corresponded to the positive inclination to that score, we defined the Ground truth cultural score for that metric as

\begin{equation}
    \textrm{GT}(d_{i}) = \frac{P}{N}
\end{equation}

\subsection{Cultural similarity score}
Equations 1 and 2 served as the basic building blocks for evaluating the LLM cultural barrier. Along with these equations, we also defined a similarity score between the LLM score ($\textrm{LLM}(d_{i})$) and the ground truth score ($\textrm{GD}(d_{i})$) for a given cultural dimension $d_{i}$, as follows \cite{wang2023cdeval} in equation 3. 

\begin{equation}
\text{Sim}(LLM, GD) = \frac{1}{1 + \sqrt{\sum_{i=1}^{6} (\textrm{LLM}(d_{i}) - \textrm{GT}(d_{i}))^2}}
\end{equation}

\subsection{Algorithm Design}
We explain the implementation and logic for evaluating the LLMs on the datasets.
We used questions from the CDEval benchmark \cite{wang2024cdevalbenchmarkmeasuringcultural}. A sample of each JSON file for each of the languages is read. We load all 6 dimensions for each language. The files are passed through the `loadOrComputeScore()` function, which checks if the scores for the dimensions of each language are computed; if the languages have been computed, we use the scores. If not, we compute the scores. We define the keys for each language, English, Malayalam, and Yoruba, and print the format for each language along with the keys for each dimension. A statement is included to get the values of the keys by using the first three items. 

In computing the scores, we take the questions from a specific language and dimension. A prompt is constructed using the question and its options for each question. The prompt is passed three times to the LLM. It then analyses the response to determine the chosen option associated with the measured cultural dimension.

The individual scores are combined using a simple average, and the cultural dimension score is computed by taking the aggregated score and passing it through the `computeCulturalDimensionScore` function for each of the dimensions of all the languages. After calculating the scores for all the dimensions, we calculate the similarity to determine how close the scores are to the ground truth.
The process is repeated through all dimensions of the languages and the results are stored in JSON files.

\begin{algorithm}[hbt!]
\caption{Algorithm Design}
\begin{algorithmic}[1]
\State \textbf{Input:} Languages \( L = \{\text{Malayalam}, \text{Yoruba}, \text{English}\} \)
\State Hofstede Dimensions \( D = \{\text{IDV}, \text{IVR}, \text{UAI}, \text{PDI}, \text{MAS}, \text{LTO}\} \)
\State JSON Directories \( F_l \) for each \( l \in L \)
\State Response Number \( R = 3 \)
\State Cache File Path
\State \textbf{Output:} Computed scores \( \hat{C}_M(l, d) \) for each language \( l \) and dimension \( d \).

\State \( M_l \gets \text{read\_json\_files}(F_l) \) for each \( l \in L \)

\For{each \( l \in L \)}
    \For{each \( d \in D \)}
        \If{\( \hat{C}_M(l, d) \) exists in cache}
            \State Retrieve \( \hat{C}_M(l, d) \) from cache
        \Else
            \State Initialize \( \text{scores} \gets [] \)
            \For{each \( q \) in \( M_l(d) \)}
                \State Construct \( \text{prompt} \) using \( q \) and options \( o_1 \) and \( o_2 \)
                \For{\( k = 1 \) to \( R \)}
                    \State \( \text{response} \gets M(\text{prompt}) \)
                    \State \( \hat{a}_k \gets \text{extract\_action}(\text{response}) \)
                    \State Append \( \hat{a}_k \) to \( \text{scores} \)
                \EndFor
            \EndFor
            \State Calculate \( \hat{C}_M(l, d) \) from \( \text{scores} \)
            \State Store \( \hat{C}_M(l, d) \) in cache
        \EndIf
    \EndFor
\EndFor

\State \textbf{Output:} the computed scores \( \hat{C}_M(l, d) \) for each \( l \) and \( d \).

\end{algorithmic}
\end{algorithm}

\section{Results}
Our study compared the cultural dimension scores obtained from GPT-4o-mini and Gemma 7B for English, Malayalam, and Yoruba languages against ground truth data. The results reveal significant disparities in the LLMs' ability to capture cultural nuances across these languages. Figures 7 through 9 represent the results of our study on the cultural awareness of large language models (LLMs) for Yoruba, Malayalam, and English. Using Hofstede's cultural dimension as a framework, these figures provide a comprehensive view of how well different LLMs do across these languages.

\begin{table}[h!]
    \centering
    \caption{Comparison of Ground Truth and Computed Scores Across Hofstede's Dimensions}
    \begin{tabularx}{0.7\textwidth}{lXXXXXX}
        \toprule
        \textbf{Language} & \textbf{PDI} & \textbf{IDV} & \textbf{MAS} & \textbf{UAI} & \textbf{LTO} & \textbf{IVR} \\
        \midrule
        \textbf{Ground Truth} & & & & & & \\
        English   & 0.40 & 0.60 & 0.62 & 0.46 & 0.50 & 0.68 \\
        Yoruba    & 0.46 & 0.48 & 0.26 & 0.41 & 0.70 & 0.36 \\
        Malayalam & 0.21 & 0.48 & 0.21 & 0.32 & 0.49 & 0.38 \\
        \midrule
        \textbf{GPT-4o-mini} & & & & & & \\
        English   & 0.83 & 0.69 & 0.81 & 0.66 & 0.31 & 0.43 \\
        Yoruba    & 0.45 & 0.29 & 0.53 & 0.29 & 0.05 & 0.13 \\
        Malayalam & 0.54 & 0.38 & 0.57 & 0.38 & 0.11 & 0.16 \\
        \midrule
        \textbf{Gemma-7b-it} & & & & & & \\
        English   & 0.84 & 0.53 & 0.80 & 0.67 & 0.17 & 0.42 \\
        Yoruba    & 0.18 & 0.20 & 0.21 & 0.15 & 0.17 & 0.27 \\
        Malayalam & 0.79 & 0.64 & 0.83 & 0.78 & 0.47 & 0.58 \\
        \bottomrule
    \end{tabularx}
    \label{tab:ground_truth_vs_computed_scores_overall}
\end{table}

\subsection{Overall Trend}
\begin{figure}[H]
    \centering
    \begin{subfigure}[b]{0.4\textwidth}
        \centering
        \includegraphics[width=\textwidth]{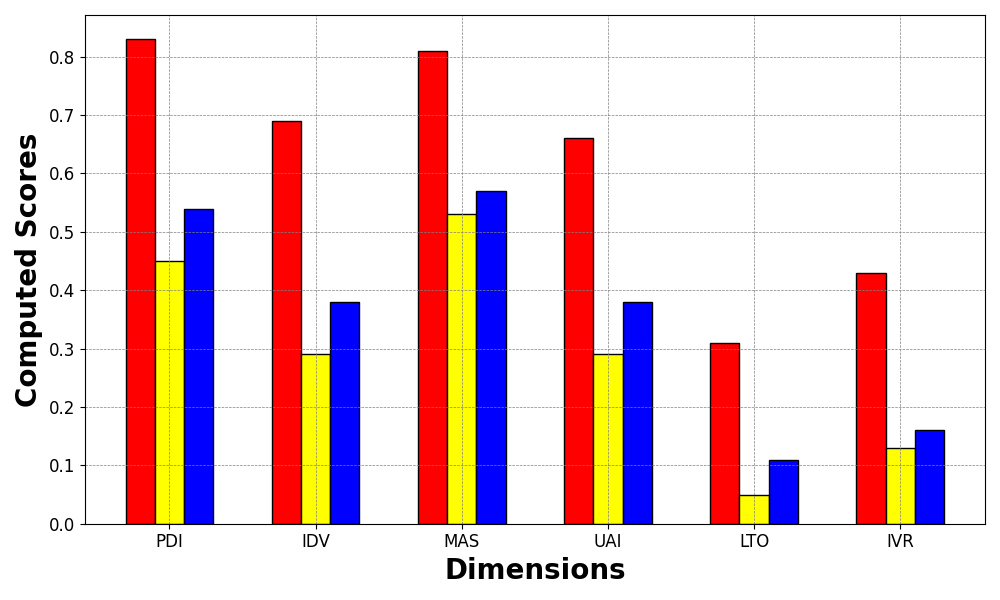}
        \caption{Computed scores using GPT-4o-mini.}
        \label{fig:gpt_scores}
    \end{subfigure}
    \hfill
    \begin{subfigure}[b]{0.4\textwidth}
        \centering
        \includegraphics[width=\textwidth]{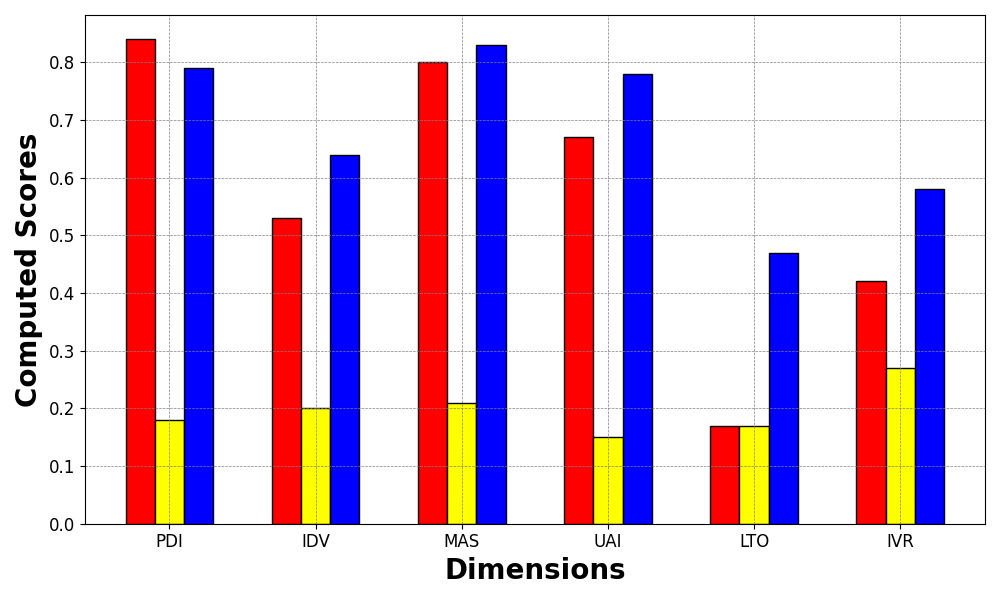}
        \caption{Computed scores using Gemma 7B.}
        \label{fig:gemma_scores}
    \end{subfigure}
    
    \vspace{1em}
    \begin{tikzpicture}
        \node[draw, fill=red, text width=2em, minimum height=1em] (english) at (0,0) {};
        \node[right=0.5cm] at (english.east) {English};
        
        \node[draw, fill=yellow, text width=2em, minimum height=1em] (yoruba) at (5,0) {};
        \node[right=0.5cm] at (yoruba.east) {Yoruba};
        
        \node[draw, fill=blue, text width=2em, minimum height=1em] (malayalam) at (10,0) {};
        \node[right=0.5cm] at (malayalam.east) {Malayalam};
    \end{tikzpicture}

    \caption{Comparison of computed scores for English, Yoruba, and Malayalam across Hofstede's 6 cultural dimensions using GPT-4o-mini and Gemma 7B.}
    \label{fig:computed_scores_comparison}
\end{figure}
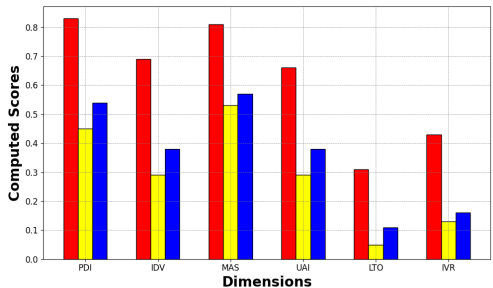
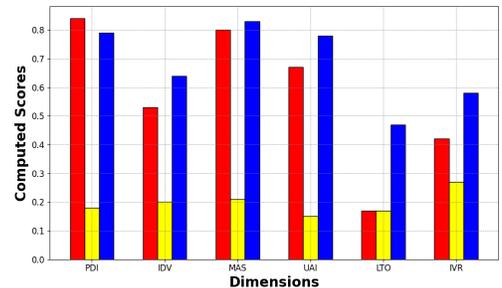

Table \ref{tab:ground_truth_vs_computed_scores_overall} compares Hofstede's Cultural Dimensions between Ground Truth and two LLMs (GPT-4o-mini and Gemma-7b-it) across English, Yoruba, and Malayalam. The computed scores for individual dimensions are plotted as bar graphs in Figure \ref{fig:computed_scores_comparison}. 

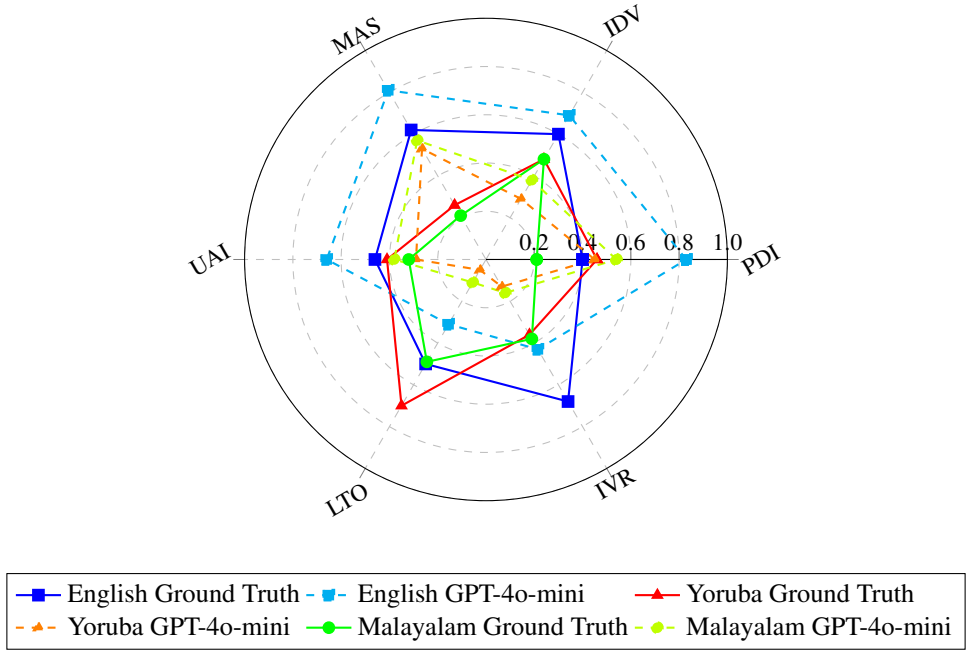
\begin{figure}[h]
    \centering
    \begin{tikzpicture}
        \begin{polaraxis}[
            width=8cm,  
            height=8cm,  
            grid=major,
            grid style={dashed,gray!50},
            tick label style={font=\small},
            ytick={0.2, 0.4, 0.6, 0.8, 1.0},
            ymin=0,
            ymax=1.0,
            xtick={0,60,120,180,240,300},
            xticklabels={
                PDI, 
                IDV, 
                MAS, 
                UAI, 
                LTO, 
                IVR
            },
            xticklabel style={font=\small, rotate=30},
            yticklabels={0.2, 0.4, 0.6, 0.8, 1.0},
            legend style={at={(0.5,-0.15)},anchor=north,legend columns=3},
            legend cell align={left}            ]
            \addplot[
                color=blue,
                mark=square*,
                mark options={fill=blue},
                thick
            ] coordinates {
                (0,0.40) (60,0.60) (120,0.62) (180,0.46) (240,0.50) (300,0.68) (360,0.40)
            };
            \addplot[
                color=cyan,
                mark=square*,
                mark options={fill=cyan},
                thick,
                dashed
            ] coordinates {
                (0,0.83) (60,0.69) (120,0.81) (180,0.66) (240,0.31) (300,0.43) (360,0.83)
            };

            \addplot[
                color=red,
                mark=triangle*,
                mark options={fill=red},
                thick
            ] coordinates {
                (0,0.46) (60,0.48) (120,0.26) (180,0.41) (240,0.70) (300,0.36) (360,0.46)
            };
            \addplot[
                color=orange,
                mark=triangle*,
                mark options={fill=orange},
                thick,
                dashed
            ] coordinates {
                (0,0.45) (60,0.29) (120,0.53) (180,0.29) (240,0.05) (300,0.13) (360,0.45)
            };

            \addplot[
                color=green,
                mark=*,
                mark options={fill=green},
                thick
            ] coordinates {
                (0,0.21) (60,0.48) (120,0.21) (180,0.32) (240,0.49) (300,0.38) (360,0.21)
            };
            \addplot[
                color=lime,
                mark=*,
                mark options={fill=lime},
                thick,
                dashed
            ] coordinates {
                (0,0.54) (60,0.38) (120,0.57) (180,0.38) (240,0.11) (300,0.16) (360,0.54)
            };

            \addlegendentry{English Ground Truth}
            \addlegendentry{English GPT-4o-mini}
            \addlegendentry{Yoruba Ground Truth}
            \addlegendentry{Yoruba GPT-4o-mini}
            \addlegendentry{Malayalam Ground Truth}
            \addlegendentry{Malayalam GPT-4o-mini}

        \end{polaraxis}
    \end{tikzpicture}

    \caption{Comparison of cultural scores for GPT-4o-mini vs Ground Truth Across Hofstede's 6 Cultural Dimensions}
    \label{fig:spider_gpt4o}
\end{figure}

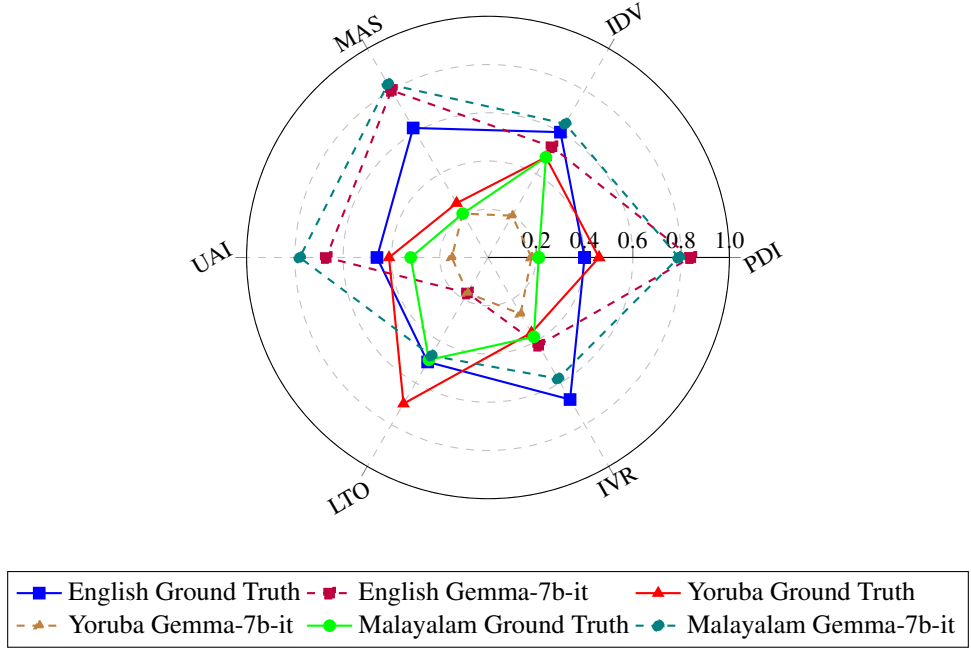
\begin{figure}[h]
    \centering
    \begin{tikzpicture}
        \begin{polaraxis}[
            width=8cm,  
            height=8cm,  
            grid=major,
            grid style={dashed,gray!50},
            tick label style={font=\small},
            ytick={0.2, 0.4, 0.6, 0.8, 1.0},
            ymin=0,
            ymax=1.0,
            xtick={0,60,120,180,240,300},
            xticklabels={
                PDI, 
                IDV, 
                MAS, 
                UAI, 
                LTO, 
                IVR
            },
            xticklabel style={font=\small, rotate=30},
            yticklabels={0.2, 0.4, 0.6, 0.8, 1.0},
            legend style={at={(0.5,-0.15)},anchor=north,legend columns=3},
            legend cell align={left}            ]
            \addplot[
                color=blue,
                mark=square*,
                mark options={fill=blue},
                thick
            ] coordinates {
                (0,0.40) (60,0.60) (120,0.62) (180,0.46) (240,0.50) (300,0.68) (360,0.40)
            };
            \addplot[
                color=purple,
                mark=square*,
                mark options={fill=purple},
                thick,
                dashed
            ] coordinates {
                (0,0.84) (60,0.53) (120,0.80) (180,0.67) (240,0.17) (300,0.42) (360,0.84)
            };

            \addplot[
                color=red,
                mark=triangle*,
                mark options={fill=red},
                thick
            ] coordinates {
                (0,0.46) (60,0.48) (120,0.26) (180,0.41) (240,0.70) (300,0.36) (360,0.46)
            };
            \addplot[
                color=brown,
                mark=triangle*,
                mark options={fill=brown},
                thick,
                dashed
            ] coordinates {
                (0,0.18) (60,0.20) (120,0.21) (180,0.15) (240,0.17) (300,0.27) (360,0.18)
            };

            \addplot[
                color=green,
                mark=*,
                mark options={fill=green},
                thick
            ] coordinates {
                (0,0.21) (60,0.48) (120,0.21) (180,0.32) (240,0.49) (300,0.38) (360,0.21)
            };
            \addplot[
                color=teal,
                mark=*,
                mark options={fill=teal},
                thick,
                dashed
            ] coordinates {
                (0,0.79) (60,0.64) (120,0.83) (180,0.78) (240,0.47) (300,0.58) (360,0.79)
            };

            \addlegendentry{English Ground Truth}
            \addlegendentry{English Gemma-7b-it}
            \addlegendentry{Yoruba Ground Truth}
            \addlegendentry{Yoruba Gemma-7b-it}
            \addlegendentry{Malayalam Ground Truth}
            \addlegendentry{Malayalam Gemma-7b-it}

        \end{polaraxis}
    \end{tikzpicture}

    \caption{Comparison of cultural scores for Gemma-7b-it vs Ground Truth Across Hofstede's 6 Cultural Dimensions}
    \label{fig:spider_gemma7b}
\end{figure}

This variation is particularly evident in dimensions like \textbf{Power Distance Index (PDI)} and \textbf{Masculinity (MAS)}, where the models tend to overestimate the cultural acceptance of hierarchy and assertiveness in these languages compared to their ground truth scores. For example, both models significantly overestimate PDI for English (Ground truth: 0.40, GPT-4o-mini: 0.83, Gemma-7b-it: 0.84) and Malayalam (Ground truth: 0.21, GPT-4o-mini: 0.54, Gemma-7b-it: 0.79). This suggests that the models may overestimate how much power inequality is tolerated in English-speaking cultures and may project those learned cultural norms onto underrepresented languages like Malayalam, where equality may be more common. For Yoruba, GPT-4o-mini closely matches the ground truth (0.46 vs 0.45), while Gemma-7b-it significantly underestimates it (0.18).

In \textbf{Individualism (IDV)}, both models slightly overestimate for English, with GPT-4o-mini being closer to the ground truth (Ground truth: 0.60, GPT-4o-mini: 0.69, Gemma-7b-it: 0.53). For Yoruba and Malayalam, both models underestimate IDV, with Gemma-7b-it showing a higher score for Malayalam (0.64) compared to the ground truth (0.48). This underestimates the communist tendencies in Yoruba and Malayalam. This underestimation can be attributed to the models being trained predominantly on Western-centric data, where individualism is more pronounced, leading to weaker performance when capturing the stronger community orientation found in these regional languages.

Both models overestimate this dimension for English in the \textbf{Uncertainty Avoidance Index (UAI)}. For Yoruba and Malayalam, GPT-4o-mini underestimates UAI, while Gemma-7b-it underestimates for Yoruba but significantly overestimates for Malayalam. This just shows that both models have trouble understanding how the culture tends to avoid risks and deal with uncertainty. The most significant differences are observed in the \textbf{Long Term Orientation (LTO)} and \textbf{Indulgence (IVR)} dimensions. Both models consistently underestimate these dimensions across all three languages, with the underestimation being particularly severe for Yoruba and Malayalam in the LTO dimension. This suggests the models struggle to capture fundamental cultural values related to future orientation, tradition, and self-restraint in these cultures.

These results indicate that while the models show some understanding of cultural dimensions, they struggle to accurately represent the nuances across different languages, particularly for non-English languages like Yoruba and Malayalam. The consistent overestimation of some dimensions (like MAS) and underestimation of others (like LTO and IVR) across languages suggest systematic biases in how these models represent cultural attributes. Figure \ref{fig:spider_gpt4o} and \ref{fig:spider_gemma7b} display the overall trend as a spider plot.

\subsection{Model's general performance across languages.}
Here, we examine the cultural sensitivity of the models across the selected languages: Yoruba, Malayalam, and English. We focus on evaluating the capabilities of GPT-4o mini and Gemma-7b-it in this regard. Equation 3 shows the cultural similarity score calculation between a model and the corresponding language's country. Table \ref{tab:similarity_scores_simplified} shows the similarity scores for both models across the three languages. 

\begin{figure}[H]
    \centering
    \begin{subfigure}[b]{0.45\textwidth}
        \centering
        \includegraphics[width=\textwidth]{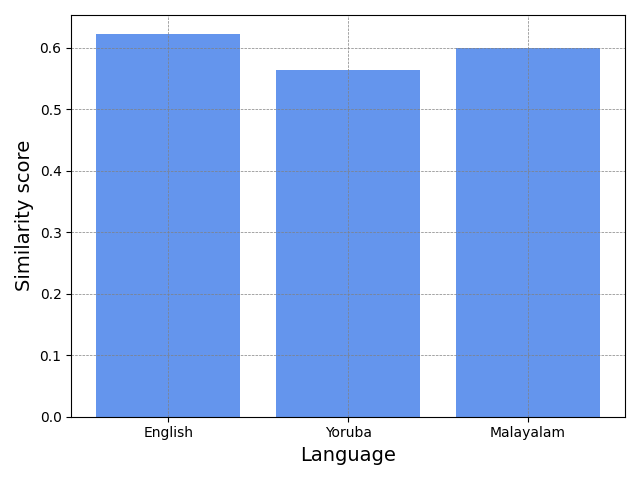}
        \caption{Similarity Scores for GPT-4o-mini for English, Yoruba, and Malayalam. As hypothesized, the LLM captures cultural nuances in English better than Yoruba and Malayalam.}
        \label{fig:similarity_score_gpt}
    \end{subfigure}
    \hfill
    \begin{subfigure}[b]{0.45\textwidth}
        \centering
        \includegraphics[width=\textwidth]{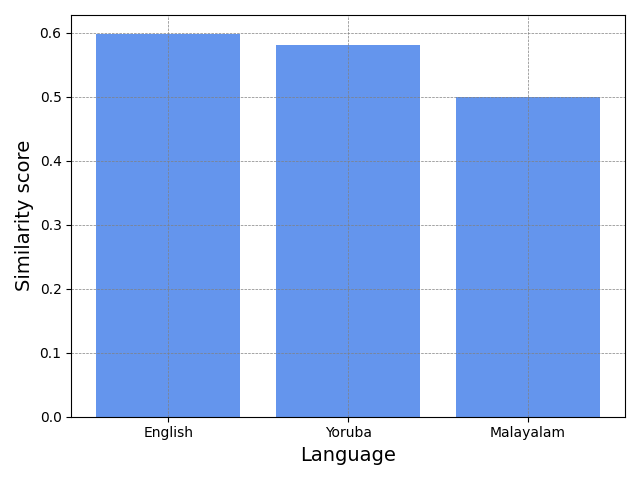}
        \caption{Similarity Scores for Gemma 7B for English, Yoruba, and Malayalam. Similar trends are observed in cultural nuance capturing, with differences in accuracy across languages.}
        \label{fig:similarity_score_gemma}
    \end{subfigure}
    
    \caption{Comparison of Similarity Scores for GPT-4o-mini and Gemma 7B for English, Yoruba, and Malayalam.}
    \label{fig:similarity_score_comparison}
\end{figure}

GPT-4o-mini performs best in English, with a similarity score of 0.6225. Malayalam was the second-highest, with a score of 0.5998, suggesting a slightly lower performance than English. On the other hand, Yoruba had the lowest score, 0.5637, indicating the model struggles most with this language. Interestingly, Gemma-7b-it shows a different pattern. It performs best with Yoruba (0.5819), followed closely by English (0.5981), while it struggles the most with Malayalam (0.5002). The results are also plotted in the bar plot of Figure \ref{fig:similarity_score_comparison} for better visualization.

\begin{table}[h!]
    \centering
    \caption{Similarity Scores Across Languages for GPT-4o-mini and Gemma-7b-it}
    \begin{tabularx}{0.6\textwidth}{X X X}
        \toprule
        \textbf{Language} & \textbf{GPT-4o-mini} & \textbf{Gemma-7b-it} \\
        \midrule
        English      & 0.6225 & 0.5981 \\
        Yoruba       & 0.5637 & 0.5819 \\
        Malayalam    & 0.5998 & 0.5002 \\
        \bottomrule
    \end{tabularx}
    \label{tab:similarity_scores_simplified}
\end{table}

The model shows a range of clustering scores from approximately 0.56 to 0.62, indicating that the model performs better than random chance (0.5) across all languages. However, this narrow range might indicate a ceiling effect in the model's ability to capture each language's cultural nuances and unique linguistic features, pointing to architectural or training limitations. The higher score for English in GPT-4o-mini is not surprising because most LLMs are trained predominantly on English data. However, the model's performance in Malayalam is notably close to English, which was unexpected but interesting. For Gemma-7b-it, the surprisingly good performance in Yoruba and the relatively poor performance in Malayalam raises questions about the model's training data and architecture.

This narrow range could imply that the model lacks a deep understanding of diverse cultural contexts. The findings also highlight that no model performs consistently well across all cultural dimensions, particularly for non-English languages. This calls for more diverse and representative training data to improve its ability to adapt to a language's unique features. 

\subsection{Performance Analysis}
Table \ref{tab:gpt4-mini} and \ref{tab:gemma-7b} show the similarity scores for GPT-4o- mini and Gemma-7b-it across the three languages. These scores are calculated based on the absolute difference between the ground truth and the GPT-4o-mini predicted scores, normalized to a 0-1 scale where 1 indicates perfect similarity. These scores reveal varying levels of accuracy across languages and dimensions. 

\begin{table}[h!]
    \centering
    \caption{Similarity Scores Across Dimensions for GPT-4o-mini and Gemma-7b-it}
    \label{fig:similarity_score_breakdown}
    \begin{subtable}[t]{0.48\textwidth}
        \centering
        \caption{GPT-4o-mini}
        \label{tab:gpt4-mini}
        \begin{tabularx}{\textwidth}{lXXX}
            \toprule
            \textbf{Dim.} & \textbf{Eng.} & \textbf{Yor.} & \textbf{Mal.} \\ \midrule
            PDI & 0.57 & 0.99 & 0.33 \\
            IDV & 0.91 & 0.81 & 0.90 \\
            MAS & 0.81 & 0.27 & 0.36 \\
            UAI & 0.80 & 0.88 & 0.94 \\
            LTO & 0.81 & 0.35 & 0.62 \\
            IVR & 0.75 & 0.77 & 0.78 \\
            \bottomrule
        \end{tabularx}
    \end{subtable}
    \hfill
    \begin{subtable}[t]{0.48\textwidth}
        \centering
        \caption{Gemma-7b-it}
        \label{tab:gemma-7b}
        \begin{tabularx}{\textwidth}{lXXX}
            \toprule
            \textbf{Dim.} & \textbf{Eng.} & \textbf{Yor.} & \textbf{Mal.} \\ \midrule
            PDI & 0.84 & 0.18 & 0.79 \\
            IDV & 0.53 & 0.20 & 0.64 \\
            MAS & 0.80 & 0.21 & 0.83 \\
            UAI & 0.67 & 0.15 & 0.78 \\
            LTO & 0.17 & 0.17 & 0.47 \\
            IVR & 0.42 & 0.27 & 0.58 \\
            \bottomrule
        \end{tabularx}
    \end{subtable}
\end{table}

English shows consistently high similarity across all dimensions in GPT-4o-mini, with IDV being the closest match (0.91). Yoruba demonstrates high similarity in PDI (0.99) and UAI (0.88) but very low similarity in MAS (0.27) and LTO (0.35). Malayalam exhibits high similarity in UAI (0.94) and IDV (0.90) but low similarity in PDI (0.33) and MAS (0.36). However, in Gemma-7b-it, English shows high similarity in PDI (0.84) and MAS (0.80) but lower similarity in LTO (0.17). Yoruba exhibits low similarity across most dimensions, particularly in PDI (0.18) and UAI (0.15), with modest similarity in IVR (0.27). Malayalam demonstrates high similarity in MAS (0.83) and UAI (0.78) but a lower similarity in LTO (0.47) and IDV (0.64).

\subsubsection*{GPT-4o-mini Performance Analysis}

\begin{table}[ht]
\centering
\begin{tabular}{l p{4cm} p{4cm} p{4cm}}
\toprule
\textbf{Dimension} & \makecell{\textbf{English} \\ \textbf{(GT vs GPT-4o)}} & \makecell{\textbf{Yoruba} \\ \textbf{(GT vs GPT-4o)}} & \makecell{\textbf{Malayalam} \\ \textbf{(GT vs GPT-4o)}} \\ 
\midrule
PDI & 0.40 vs 0.83 \,(-0.43) & 0.46 vs 0.45 \,(0.01) & 0.21 vs 0.54 \,(-0.33) \\
IDV & 0.60 vs 0.69 \,(-0.09) & 0.48 vs 0.29 \,(0.19) & 0.48 vs 0.38 \,(0.10) \\
MAS & 0.62 vs 0.81 \,(-0.19) & 0.26 vs 0.53 \,(-0.27) & 0.21 vs 0.57 \,(-0.36) \\
UAI & 0.46 vs 0.66 \,(-0.20) & 0.41 vs 0.29 \,(0.12) & 0.32 vs 0.38 \,(-0.06) \\
LTO & 0.50 vs 0.31 \,(0.19) & 0.70 vs 0.05 \,(0.65) & 0.49 vs 0.11 \,(0.38) \\
IVR & 0.68 vs 0.43 \,(0.25) & 0.36 vs 0.13 \,(0.23) & 0.38 vs 0.16 \,(0.22) \\
\bottomrule
\label{gpt4o}

\end{tabular}
\caption{Comparison of Ground Truth (GT) and GPT-4o-mini's predictions for each dimension across three languages. Positive values in parentheses indicate underestimation, while negative values indicate overestimation by GPT-4o-mini.}
\label{tab:performance}
\end{table}

GPT-4o-mini tends to overestimate most dimensions for English, especially PDI and MAS, as shown in Table 5. For Yoruba, it shows mixed results, underestimating some dimensions (IDV, UAI, LTO, IVR) and overestimating others (MAS). The LTO dimension shows the most significant discrepancy for Yoruba. The model overestimates PDI, MAS, and UAI in Malayalam while underestimating IDV, LTO, and IVR.

\subsubsection*{Gemma-7b-it Performance Analysis}

\begin{table}[ht]
\centering
\begin{tabular}{l p{4cm} p{4cm} p{4cm}}
\toprule
\textbf{Dimension} & \makecell{\textbf{English} \\ \textbf{(GT vs Gemma)}} & \makecell{\textbf{Yoruba} \\ \textbf{(GT vs Gemma)}} & \makecell{\textbf{Malayalam} \\ \textbf{(GT vs Gemma)}} \\
\midrule
PDI & 0.40 vs 0.84 \,(-0.44) & 0.46 vs 0.18 \,(0.28) & 0.21 vs 0.79 \,(-0.58) \\
IDV & 0.60 vs 0.53 \,(0.07)  & 0.48 vs 0.20 \,(0.28) & 0.48 vs 0.64 \,(-0.16) \\
MAS & 0.62 vs 0.80 \,(-0.18) & 0.26 vs 0.21 \,(0.05) & 0.21 vs 0.83 \,(-0.62) \\
UAI & 0.46 vs 0.67 \,(-0.21) & 0.41 vs 0.15 \,(0.26) & 0.32 vs 0.78 \,(-0.46) \\
LTO & 0.50 vs 0.17 \,(0.33)  & 0.70 vs 0.17 \,(0.53) & 0.49 vs 0.47 \,(0.02) \\
IVR & 0.68 vs 0.42 \,(0.26)  & 0.36 vs 0.27 \,(0.09) & 0.38 vs 0.58 \,(-0.20) \\
\bottomrule
\end{tabular}
\caption{Comparison of Ground Truth (GT) and Gemma's predictions for each dimension across three languages. Positive values in parentheses indicate underestimation, while negative values indicate overestimation by Gemma.}
\end{table}

Gemma-7b-it also tends to overestimate most dimensions for English, particularly PDI and MAS. For Yoruba, it consistently underestimates most dimensions, with only LTO showing the most significant discrepancy. In Malayalam, Gemma-7b-it significantly overestimates most dimensions, with very large discrepancies in PDI and MAS. 


Both models show similar trends in \textbf{English}, overestimating most dimensions, especially PDI and MAS. GPT-4o-mini shows slightly better overall similarity scores. In \textbf{Yoruba}, the models perform differently. While GPT-4o-mini shows mixed results with high accuracy in some dimensions (PDI, UAI), Gemma-7b-it consistently underestimates most dimensions. 
GPT-4o-mini's predictions for Yoruba show higher overall similarity to ground truth. Both models struggle with accurate predictions in \textbf{Malayalam}. GPT-4o-mini shows varied performance across dimensions, while Gemma-7b-it significantly overestimates most dimensions. GPT-4o-mini generally shows higher similarity scores for Malayalam, as shown in Table \ref{tab:similarity_scores_simplified}.

These results highlight the complex nature of predicting cultural dimensions across different languages. While both models show promise in certain areas, they also demonstrate significant challenges in accurately capturing nuanced cultural aspects, particularly for non-English languages. The varying performance across dimensions and languages suggests that further refinement is needed to improve these models' understanding of diverse cultural contexts.

\subsection{Discussion of Model Biases and Performance}
The discrepancies between the cultural dimension scores predicted by the models and the ground truth for Yoruba and Malayalam suggest that underlying model biases, particularly those related to the training data, play a significant role. Large language models are predominantly trained on English-centric data, with minimal representation of regional languages like Yoruba and Malayalam. The Common Crawl data \ref{tab:language_distribution} shows that English dominates LLM training data ($45\%$), while Yoruba and Malayalam account for only $0.008\%$ and $0.0226\%$, respectively. This stark imbalance in the training datasets contributes to the models' over- or underestimation of cultural dimensions for these languages.

These biases arise because models are exposed to a disproportionate amount of content from English-speaking cultures. This leads to internalizing cultural norms that may not align with those in underrepresented regions. For instance, the overestimation of Power Distance (PDI) in Malayalam (ground truth: 0.21, GPT-4o-mini: 0.54) likely stems from hierarchical assumptions in English-speaking cultures less relevant in Malayalam-speaking contexts. Similarly, the underestimation of Long-Term Orientation (LTO) for Yoruba (ground truth: 0.70, GPT-4o-mini: 0.05) indicates the models' limited understanding of future-oriented cultural values in West African societies.

The differences between non-English languages and English present challenges for language models. Yoruba and Malayalam have unique syntactic and grammatical structures that differ from English, which can lead to misunderstandings. The lack of cultural contexts in training data exacerbates these challenges. We need a more balanced dataset that includes regional languages to address this. The imbalance in training data impacts the models' linguistic abilities and hinders their adaptability to different cultures. This restricts their usefulness in multilingual and multicultural settings. Models perform better in English, indicating the need for improved performance in languages like Yoruba and Malayalam. This requires larger, diverse datasets and adaptable model architectures to accommodate linguistic and cultural diversity better.

\section{Conclusion and Discussion}
In this study, our main goal was to understand whether Large Language Models (LLMs) are able to effectively capture the cultural nuances of regional languages. To achieve this goal, we looked at 2 languages: Malayalam (spoken by $\approx$ 38 million people in the state of Kerala, India) and Yoruba (spoken by $\approx$ 45 million people largely concentrated in West Africa). We selected these 2 languages because in terms of native speakers, these languages represent a minority - only 1-2 $\%$ of the entire population outside Africa speaks Yoruba, and only 1-2 $\%$ of the entire population outside India speaks Malayalam. As much as $\approx$ 60 $\%$ of the training data for all modern LLMs is based on the English language. This would naturally lead to LLMs having a poor understanding and poor functionality for regional languages and cultures. To demonstrate this lack of LLM understanding of non-English languages, we wanted to choose languages originating from areas that are not English-centric. People speaking Malayalam or Yoruba mostly rely on their local language for communication, entertainment, and employment. That's another major reason why we selected these 2 languages. \newline

To quantify the cultural awareness of LLM-based responses, we looked at 6 cultural metrics, also called the Hofstede's metrics: Power Distance (PDI), Individualism (IDV), Motivation towards Achievement and Success (MAS), Uncertainty Avoidance (UAV), Long Term Orientation (LTO) and Indulgence (IVR). For each of these metrics, we designed questions and recorded the LLM responses. The LLM responses helped us quantify the LLM cultural score for each metric. We then compared the LLM cultural scores with the ground truth data for each of the 6 metrics, and obtained a cultural similarity score. The cultural similarity score is a composite score that includes PDI, IDV, MAS, UAV, LTO, and IVR. It indicates the ability of the LLM to effectively capture the nuances of the regional languages. A higher score indicates that the LLM better understands the regional language than a lower score. \newline

Our results indicate that the cultural similarity score obtained for the languages of Malayalam and Yoruba is significantly lower than English, for all the LLMs we tested: Open AI's GPT-4o-mini and Gemma-7b-it. This clearly demonstrates that all modern LLMs are vastly inadequate in their understanding and, consequently, their functionality for regional languages. As the influence of LLMs grows more powerful, the English-centric bias in LLMs has the potential to significantly affect the lives of people speaking regional languages, further highlighting the importance of this study. \newline

\subsection{Key Findings}
Our results indicate that the cultural similarity score obtained for the languages of Malayalam and Yoruba is significantly lower than English for all the LLMs we tested: Open AI's GPT-4o-mini and Gemma-7b-it. This clearly demonstrates that all modern LLMs are vastly inadequate in their understanding and, consequently, their functionality for regional languages. As the influence of LLMs grows more powerful, the English-centric bias in LLMs has the potential to significantly affect the lives of people speaking regional languages, further highlighting the importance of this study \cite{blodgett-etal-2020-language}.

\subsection{Implications for Cultural Sensitivity}
Our findings underscore the critical need for improved cultural sensitivity in LLMs, particularly for underrepresented languages. The significant disparity in cultural similarity scores between English and regional languages like Malayalam and Yoruba highlights a potential for misunderstanding and misrepresentation of these cultures in AI-driven applications. This lack of cultural awareness in LLMs could lead to inappropriate or ineffective interactions in various domains, including education, healthcare, and local governance \cite{paullada2020data}. As LLMs become increasingly integrated into global communication systems, addressing this cultural gap is crucial.

\subsection{Ethical Considerations}
The observed performance gap in cultural awareness between English and underrepresented languages raises important ethical considerations. This disparity could potentially perpetuate and exacerbate existing inequalities in access to and benefit from AI technologies, potentially leading to a form of technological colonialism if not addressed \cite{Sambasivan2021EveryoneWT}. It is imperative that the development of LLMs takes into account the diverse cultural landscapes they are intended to serve, ensuring fair representation and understanding of all languages and cultures.

\subsection{Limitations}
One of the primary challenges and limitations of the present study was collecting ground truth data for Malayalam and Yoruba. While robust documentation of human cultural dimension scores from Hofstede's survey exists for major global languages spoken in countries like the United States, Germany, and China, Hofstede's cultural survey information for Malayalam and Yoruba is very limited. To bridge this gap, we constructed our survey of questions, managing to collect approximately 100 questions for both languages. However, this limited sample size may not fully capture the cultural nuances of these languages.
Additionally, the use of machine translation (Google Translate API) to translate the English questions into Malayalam and Yoruba may have introduced inaccuracies or cultural misinterpretations. The binary (Yes/No) nature of the questions may also oversimplify complex cultural concepts.
Furthermore, our study focused on only two regional languages, which, while providing valuable insights, may not be fully representative of the vast linguistic diversity in the world. The performance of LLMs in other underrepresented languages may vary.
\subsection{Future Work}
Our research opens up several avenues for future work:
\begin{enumerate}
\item \textbf{Expanding the dataset:}
Collecting more responses and making this growing dataset public is one of our long-term objectives. This will provide a more comprehensive ground truth for Malayalam and Yoruba cultural dimensions.
\item \textbf{Extending to more languages:} 
Although we looked at only two regional languages, the applications of this study extend to more than 1000 regional languages worldwide. Future research should include a wider range of underrepresented languages \cite{joshi-etal-2020-state}.

\item \textbf{Improving evaluation methods:} 
Developing more nuanced evaluation techniques that go beyond binary questions could provide a more accurate assessment of LLMs' cultural awareness. This could include:
\begin{itemize}
    \item Incorporating context-dependent scenarios to assess cultural understanding in specific situations.
    \item Developing multi-dimensional scoring systems that capture the complexity of cultural interactions.
    \item Utilizing natural language generation tasks to evaluate the LLM's ability to produce culturally appropriate responses.
\end{itemize}

\item \textbf{Exploring alternative cultural frameworks:} 
While Hofstede's dimensions provide a valuable starting point, future work could explore other cultural frameworks or develop new ones specifically tailored for AI evaluation \cite{masoud2024culturalalignmentlargelanguage}.

\item \textbf{Investigating the impact of fine-tuning:} 
Future studies could explore how fine-tuning LLMs on culturally diverse datasets affects their cultural awareness and performance across different languages.
\end{enumerate}

Although we looked at only 2 regional languages, the applications of this study extend to more than the 1000 regional languages of the world. As LLMs become more powerful, we need to make sure that regional languages are not getting left behind, to ensure an equitable world for all. Our study highlights the urgent need for more diverse and culturally enriched training datasets for LLMs to improve their performance in underrepresented languages.
The results of this study emphasize the importance of addressing the cultural awareness gap in LLMs for the thousands of regional languages of the world. As LLMs become increasingly influential, ensuring equitable representation and understanding of regional languages is crucial for creating a more inclusive global AI landscape.

\section{Github}
Here is the GitHub repo where we have hosted our code: \href{https://github.com/fiifidawson/LLM-Cultural-Awareness}{LLM-Cultural-Awareness}

\bibliographystyle{unsrt}  
\bibliography{references}

\begin{thebibliography}{10}

\bibitem{naveed2024comprehensiveoverviewlargelanguage}
Humza Naveed, Asad~Ullah Khan, Shi Qiu, Muhammad Saqib, Saeed Anwar, Muhammad Usman, Naveed Akhtar, Nick Barnes, and Ajmal Mian.
\newblock A comprehensive overview of large language models, 2024.

\bibitem{fan_li_ma_lee_yu_hemphill_2017}
Lizhou Fan, Lingyao Li, Zihui Ma, Sanggyu Lee, Huizi Yu, and Libby Hemphill.
\newblock A bibliometric review of large language models research from 2017 to 2023, 2017.

\bibitem{lai2024llmsenglishscalingmultilingual}
Wen Lai, Mohsen Mesgar, and Alexander Fraser.
\newblock Llms beyond english: Scaling the multilingual capability of llms with cross-lingual feedback, 2024.

\bibitem{commoncrawl}
Common Crawl.
\newblock Common crawl language statistics.
\newblock \url{https://commoncrawl.github.io/cc-crawl-statistics/plots/languages}, 2024.
\newblock [Accessed: 2024-08-01].

\bibitem{blodgett-etal-2020-language}
Su~Lin Blodgett, Solon Barocas, Hal Daum{\'e}~III, and Hanna Wallach.
\newblock Language (technology) is power: A critical survey of {``}bias{''} in {NLP}.
\newblock In Dan Jurafsky, Joyce Chai, Natalie Schluter, and Joel Tetreault, editors, {\em Proceedings of the 58th Annual Meeting of the Association for Computational Linguistics}, pages 5454--5476, Online, July 2020. Association for Computational Linguistics.

\bibitem{vera_neplenbroek__2024}
Vera Neplenbroek, Arianna Bisazza, and Raquel Fernández.
\newblock Mbbq: A dataset for cross-lingual comparison of stereotypes in generative llms, 2024.

\bibitem{Navigli2023BiasesIL}
Roberto Navigli, Simone Conia, and Bj{\"o}rn Ross.
\newblock Biases in large language models: Origins, inventory, and discussion.
\newblock {\em ACM Journal of Data and Information Quality}, 15:1 -- 21, 2023.

\bibitem{shafayat2024multifactassessingmultilingualllms}
Sheikh Shafayat, Eunsu Kim, Juhyun Oh, and Alice Oh.
\newblock Multi-fact: Assessing multilingual llms' multi-regional knowledge using factscore, 2024.

\bibitem{tanja_samardzic__2024}
Tanja Samardzic, Ximena Gutierrez-Vasques, Christian Bentz, Steven Moran, and Olga Pelloni.
\newblock A measure for transparent comparison of linguistic diversity in multilingual nlp data sets.
\newblock {\em arXiv.org}, abs/2403.03909, 2024.

\bibitem{tarek_naous__2023}
Tarek Naous, Michael~J. Ryan, and Wei Xu.
\newblock Having beer after prayer? measuring cultural bias in large language models.
\newblock {\em arXiv.org}, abs/2305.14456, 2023.

\bibitem{levy2023comparingbiasesimpactmultilingual}
Sharon Levy, Neha~Anna John, Ling Liu, Yogarshi Vyas, Jie Ma, Yoshinari Fujinuma, Miguel Ballesteros, Vittorio Castelli, and Dan Roth.
\newblock Comparing biases and the impact of multilingual training across multiple languages, 2023.

\bibitem{ochieng2024metricsevaluatingllmseffectiveness}
Millicent Ochieng, Varun Gumma, Sunayana Sitaram, Jindong Wang, Vishrav Chaudhary, Keshet Ronen, Kalika Bali, and Jacki O'Neill.
\newblock Beyond metrics: Evaluating llms' effectiveness in culturally nuanced, low-resource real-world scenarios, 2024.

\bibitem{laurel_guthrie__2023}
Laurel Guthrie, Joseph Mkandawire, E.~Anna Stevenson, Sharon Bonya, Brent Sherwin, Moses Kasumba, Linda Hong, Yevgeniya Ioffe, and Sharon~S. Lum.
\newblock Lessons in cultural adaptations: Translation of european organization for research and treatment of cancer quality of life questionnaire cervical cancer module from english to chichewa in malawi.
\newblock {\em Journal of Surgical Research}, 2023.

\bibitem{m__zaki__2024}
M.~Zaki and Umar Saeed.
\newblock Bridging linguistic divides: The impact of ai-powered translation systems on communication equity and inclusion.
\newblock {\em Journal of translation and language studies}, 2024.

\bibitem{razvan_cristian_voicu__2024}
Razvan~Cristian Voicu, Aarush~Kaunteya Pande, M.~Hassan Tanveer, and Yusun Chang.
\newblock Communication interchange for artificial intelligence systems.
\newblock 2024.

\bibitem{10.1145/3442188.3445922}
Emily~M. Bender, Timnit Gebru, Angelina McMillan-Major, and Shmargaret Shmitchell.
\newblock On the dangers of stochastic parrots: Can language models be too big?
\newblock In {\em Proceedings of the 2021 ACM Conference on Fairness, Accountability, and Transparency}, FAccT '21, page 610–623, New York, NY, USA, 2021. Association for Computing Machinery.

\bibitem{toreini2019relationshiptrustaitrustworthy}
Ehsan Toreini, Mhairi Aitken, Kovila Coopamootoo, Karen Elliott, Carlos~Gonzalez Zelaya, and Aad van Moorsel.
\newblock The relationship between trust in ai and trustworthy machine learning technologies, 2019.

\bibitem{xunhui_yuan__2024}
Xunhui Yuan, Jinglu Hu, and Qian Zhang.
\newblock A comparative analysis of cultural alignment in large language models in bilingual contexts, 2024.

\bibitem{pedro_jos_posada_gmez_2023}
Pedro José~Posada Gómez.
\newblock Towards a praxis for intercultural ethics in explainable ai, 2023.

\bibitem{ao_xiang_2022}
Ao~Xiang.
\newblock Diversity must be at the heart of equitable ai development.
\newblock 2022.

\bibitem{peter_brien_2022}
Peter Brien.
\newblock Diversity and inclusion in artificial intelligence.
\newblock 2022.

\bibitem{Sambasivan2021EveryoneWT}
Nithya Sambasivan, Shivani Kapania, Hannah Highfill, Diana Akrong, Praveen~K. Paritosh, and Lora Aroyo.
\newblock “everyone wants to do the model work, not the data work”: Data cascades in high-stakes ai.
\newblock {\em Proceedings of the 2021 CHI Conference on Human Factors in Computing Systems}, 2021.

\bibitem{an_analytics_of_culture_modeling_subjectivity_scalability__contextuality_and_temporality_2022}
An analytics of culture: Modeling subjectivity, scalability, contextuality, and temporality, 2022.

\bibitem{kharchenko2024well}
Julia Kharchenko, Tanya Roosta, Aman Chadha, and Chirag Shah.
\newblock How well do llms represent values across cultures? empirical analysis of llm responses based on hofstede cultural dimensions.
\newblock {\em arXiv preprint arXiv:2406.14805}, 2024.

\bibitem{herskovits1955cultural}
Melville~J Herskovits.
\newblock Cultural anthropology.
\newblock 1955.

\bibitem{hall2003sociology}
John~R Hall, Mary~Jo Neitz, and Marshall Battani.
\newblock {\em Sociology on culture}.
\newblock Psychology Press, 2003.

\bibitem{hovy-spruit-2016-social}
Dirk Hovy and Shannon~L. Spruit.
\newblock The social impact of natural language processing.
\newblock In Katrin Erk and Noah~A. Smith, editors, {\em Proceedings of the 54th Annual Meeting of the Association for Computational Linguistics (Volume 2: Short Papers)}, pages 591--598, Berlin, Germany, August 2016. Association for Computational Linguistics.

\bibitem{tricia_l__merkley__2022}
Tricia~L. Merkley, Carrie Esopenko, Vanessa Zizak, Robert~M. Bilder, Adriana~M. Strutt, David~F. Tate, and Andrei Irimia.
\newblock Challenges and opportunities for harmonization of cross-cultural neuropsychological data.
\newblock {\em Neuropsychology (journal)}, 2022.

\bibitem{joshi-etal-2020-state}
Pratik Joshi, Sebastin Santy, Amar Budhiraja, Kalika Bali, and Monojit Choudhury.
\newblock The state and fate of linguistic diversity and inclusion in the {NLP} world.
\newblock In Dan Jurafsky, Joyce Chai, Natalie Schluter, and Joel Tetreault, editors, {\em Proceedings of the 58th Annual Meeting of the Association for Computational Linguistics}, pages 6282--6293, Online, July 2020. Association for Computational Linguistics.

\bibitem{lu-ng-2021-conundrums}
Jing Lu and Vincent Ng.
\newblock Conundrums in event coreference resolution: Making sense of the state of the art.
\newblock In Marie-Francine Moens, Xuanjing Huang, Lucia Specia, and Scott Wen-tau Yih, editors, {\em Proceedings of the 2021 Conference on Empirical Methods in Natural Language Processing}, pages 1368--1380, Online and Punta Cana, Dominican Republic, November 2021. Association for Computational Linguistics.

\bibitem{kreutzer-etal-2022-quality}
Julia Kreutzer, Isaac Caswell, Lisa Wang, Ahsan Wahab, Daan van Esch, Nasanbayar Ulzii-Orshikh, Allahsera Tapo, Nishant Subramani, Artem Sokolov, Claytone Sikasote, Monang Setyawan, Supheakmungkol Sarin, Sokhar Samb, Beno{\^\i}t Sagot, Clara Rivera, Annette Rios, Isabel Papadimitriou, Salomey Osei, Pedro~Ortiz Suarez, Iroro Orife, Kelechi Ogueji, Andre~Niyongabo Rubungo, Toan~Q. Nguyen, Mathias M{\"u}ller, Andr{\'e} M{\"u}ller, Shamsuddeen~Hassan Muhammad, Nanda Muhammad, Ayanda Mnyakeni, Jamshidbek Mirzakhalov, Tapiwanashe Matangira, Colin Leong, Nze Lawson, Sneha Kudugunta, Yacine Jernite, Mathias Jenny, Orhan Firat, Bonaventure F.~P. Dossou, Sakhile Dlamini, Nisansa de~Silva, Sakine {\c{C}}abuk~Ball{\i}, Stella Biderman, Alessia Battisti, Ahmed Baruwa, Ankur Bapna, Pallavi Baljekar, Israel~Abebe Azime, Ayodele Awokoya, Duygu Ataman, Orevaoghene Ahia, Oghenefego Ahia, Sweta Agrawal, and Mofetoluwa Adeyemi.
\newblock Quality at a glance: An audit of web-crawled multilingual datasets.
\newblock {\em Transactions of the Association for Computational Linguistics}, 10:50--72, 2022.

\bibitem{adilazuarda2024measuringmodelingculturellms}
Muhammad~Farid Adilazuarda, Sagnik Mukherjee, Pradhyumna Lavania, Siddhant Singh, Alham~Fikri Aji, Jacki O'Neill, Ashutosh Modi, and Monojit Choudhury.
\newblock Towards measuring and modeling "culture" in llms: A survey, 2024.

\bibitem{chen_liu__2023}
Chen Liu, Fajri Koto, T.~E. Baldwin, and Iryna Gurevych.
\newblock Are multilingual llms culturally-diverse reasoners? an investigation into multicultural proverbs and sayings.
\newblock {\em arXiv.org}, abs/2309.08591, 2023.

\bibitem{binwei_yao__2023}
Binwei Yao, Ming Jiang, Diyi Yang, and Junjie Hu.
\newblock Empowering llm-based machine translation with cultural awareness.
\newblock {\em arXiv.org}, abs/2305.14328, 2023.

\bibitem{vig2019multiscalevisualizationattentiontransformer}
Jesse Vig.
\newblock A multiscale visualization of attention in the transformer model, 2019.

\bibitem{a_mane_2024}
A~Mane.
\newblock Unlocking machine learning model decisions: A comparative analysis of lime and shap for enhanced interpretability, 2024.

\bibitem{C_fka_2023}
Ondřej Cífka and Antoine Liutkus.
\newblock Black-box language model explanation by context length probing.
\newblock In {\em Proceedings of the 61st Annual Meeting of the Association for Computational Linguistics (Volume 2: Short Papers)}, page 1067–1079. Association for Computational Linguistics, 2023.

\bibitem{wang2024cdevalbenchmarkmeasuringcultural}
Yuhang Wang, Yanxu Zhu, Chao Kong, Shuyu Wei, Xiaoyuan Yi, Xing Xie, and Jitao Sang.
\newblock Cdeval: A benchmark for measuring the cultural dimensions of large language models, 2024.

\bibitem{masoud2024culturalalignmentlargelanguage}
Reem~I. Masoud, Ziquan Liu, Martin Ferianc, Philip Treleaven, and Miguel Rodrigues.
\newblock Cultural alignment in large language models: An explanatory analysis based on hofstede's cultural dimensions, 2024.

\bibitem{myung2024blend}
Junho Myung, Nayeon Lee, Yi~Zhou, Jiho Jin, Rifki~Afina Putri, Dimosthenis Antypas, Hsuvas Borkakoty, Eunsu Kim, Carla Perez-Almendros, Abinew~Ali Ayele, et~al.
\newblock Blend: A benchmark for llms on everyday knowledge in diverse cultures and languages.
\newblock {\em arXiv preprint arXiv:2406.09948}, 2024.

\bibitem{rao2024normad}
Abhinav Rao, Akhila Yerukola, Vishwa Shah, Katharina Reinecke, and Maarten Sap.
\newblock Normad: A benchmark for measuring the cultural adaptability of large language models.
\newblock {\em arXiv preprint arXiv:2404.12464}, 2024.

\bibitem{ehsan2020humancenteredexplainableaireflective}
Upol Ehsan and Mark~O. Riedl.
\newblock Human-centered explainable ai: Towards a reflective sociotechnical approach, 2020.

\bibitem{arrieta2019explainableartificialintelligencexai}
Alejandro~Barredo Arrieta, Natalia Díaz-Rodríguez, Javier~Del Ser, Adrien Bennetot, Siham Tabik, Alberto Barbado, Salvador García, Sergio Gil-López, Daniel Molina, Richard Benjamins, Raja Chatila, and Francisco Herrera.
\newblock Explainable artificial intelligence (xai): Concepts, taxonomies, opportunities and challenges toward responsible ai, 2019.

\bibitem{guidotti2018surveymethodsexplainingblack}
Riccardo Guidotti, Anna Monreale, Salvatore Ruggieri, Franco Turini, Dino Pedreschi, and Fosca Giannotti.
\newblock A survey of methods for explaining black box models, 2018.

\bibitem{rudin2019stopexplainingblackbox}
Cynthia Rudin.
\newblock Stop explaining black box machine learning models for high stakes decisions and use interpretable models instead, 2019.

\bibitem{doshivelez2017rigorousscienceinterpretablemachine}
Finale Doshi-Velez and Been Kim.
\newblock Towards a rigorous science of interpretable machine learning, 2017.

\bibitem{doran2017doesexplainableaireally}
Derek Doran, Sarah Schulz, and Tarek~R. Besold.
\newblock What does explainable ai really mean? a new conceptualization of perspectives, 2017.

\bibitem{hofstede1984}
Geert Hofstede.
\newblock Cultural dimensions in management and planning.
\newblock {\em Asia Pacific journal of management}, 1(2):81--99, 1984.

\bibitem{hofstede2010}
Geert Hofstede, Gert~Jan Hofstede, and Michael Minkov.
\newblock {\em Cultures and organizations: Software of the mind}.
\newblock McGraw-Hill, 2010.

\bibitem{minkov2011}
Michael Minkov and Geert Hofstede.
\newblock Cross-cultural analysis: The science and art of comparing the world's modern societies and their cultures.
\newblock {\em Cross-Cultural Research}, 45(3):259--298, 2011.

\bibitem{hofstede2002}
Geert Hofstede.
\newblock Dimensions do not exist: A reply to brendan mcsweeney.
\newblock {\em Human relations}, 55(11):1355--1361, 2002.

\bibitem{hofstede2011}
Geert Hofstede.
\newblock Dimensionalizing cultures: The hofstede model in context.
\newblock {\em Online readings in psychology and culture}, 2(1):8, 2011.

\bibitem{wang2023cdeval}
Yuhang Wang, Yanxu Zhu, Chao Kong, Shuyu Wei, Xiaoyuan Yi, Xing Xie, and Jitao Sang.
\newblock Cdeval: A benchmark for measuring the cultural dimensions of large language models.
\newblock {\em arXiv preprint arXiv:2311.16421}, 2023.
\newblock \url{https://arxiv.org/pdf/2311.16421}.

\bibitem{paullada2020data}
Amandalynne Paullada, Inioluwa~Deborah Raji, Emily~M Bender, Emily Denton, and Alex Hanna.
\newblock Data and its (dis) contents: A survey of dataset development and use in machine learning research.
\newblock {\em Patterns}, 2(11):100336, 2021.

\end{thebibliography}
\end{document}